%% file: Villagrasa.tex
\documentclass[onecolumn,english,aps,showpacs,prc,superscriptaddress,preprintnumbers,floatfix,footinbib]{revtex4}
\usepackage{graphicx}
\begin{document}

\begin{center}
\linespread{1.3}

{\bf Spallation Residues in the Reaction $^{56}Fe + p$ at 0.3, 0.5, 0.75, 1.0 and 1.5 A~GeV.}

\vspace{0.5cm}
\footnotesize{
C.~VILLAGRASA-CANTON$^{a,1}$,
A.~BOUDARD$^a$,
J.-E.~DUCRET$^a$,
B.~FERNANDEZ$^a$,
S.~LERAY$^a$,
C.~VOLANT$^a$,
P.~ARMBRUSTER$^b$,
T.~ENQVIST$^b$,
F.~HAMMACHE$^b$,
K.~HELARIUTTA$^b$,
B.~JURADO$^b$,
M.-V.~RICCIARDI$^b$,
K.-H.~SCHMIDT$^b$,
K.~S\"UMMERER$^b$,
F.~VIV\`ES$^b$,
O.~YORDANOV$^b$,
L.~AUDOUIN$^c$,
C.-O.~BACRI$^c$,
L.~FERRANT$^c$,
P.~NAPOLITANI$^{b,c,2}$,
F.~REJMUND$^{c,2}$,
C.~ST\'EPHAN$^c$,
L.~TASSAN-GOT$^c$,
J.~BENLLIURE$^d$,
E.~CASAREJOS$^d$,
M.~FERNANDEZ-ORDO\~NEZ$^{d,3}$,
J.~PEREIRA$^{d,4}$,
S.~CZAJKOWSKI$^e$,
D.KARAMANIS$^e$,
M.~PRAVIKOFF$^e$,
J.S.~GEORGE$^f$,
R.A.~MEWALDT$^f$,
N.~YANASAK$^f$,
M.~WIEDENBECK$^g$,
J.J.~CONNELL$^h$,
T.~FAESTERMANN$^i$,
A.~HEINZ$^j$,
A.~JUNGHANS$^k$\\}
\end{center}

\vspace{0.5cm}
\linespread{1.}
\begin{flushleft}
\footnotesize{
$^a$DAPNIA/SPhN, CEA/Saclay, F-91191 Gif-sur-Yvette Cedex, France\\
$^b$GSI, Planckstrasse 1, D-64291 Darmstadt, Germany\\
$^c$IPN Orsay, BP 1, F-91406 Orsay Cedex, France\\
$^d$University of Santiago de Compostela, 15706 Santiago de Compostela, Spain\\
$^e$CEN Bordeaux-Gradignan, UMR 5797 CNRS/IN2P3 - Universit\'e Bordeaux 1, BP 120, F-33175, Gradignan, France\\
$^f$California Institute of Technology, Pasadena, CA 91125 USA\\
$^g$Jet Propulsion Laboratory,California Institute of Technology, Pasadena, CA 91109 USA\\
$^h$University of New Hampshire, Durham, NH 03824, USA\\
$^i$TU Munich, 85747 Garching, Germany\\
$^j$Argonne National Laboratory, Argonne, IL 60439-4083 USA\\
$^k$CENPA/University of Washington, Seattle WA 98195 USA}

\end{flushleft}

\today

\vspace{1.0cm}
\linespread{1.}

\noindent
{\bf Abstract} - The spallation residues produced in the bombardment of $^{56}Fe$ at 1.5, 1.0, 0.75, 0.5 and 0.3 A~GeV on a liquid-hydrogen target have been measured using the reverse kinematics technique and the Fragment Separator at GSI (Darmstadt). This technique has permitted the full identification in charge and mass of all isotopes produced with cross-sections larger than $10^{-2}$ mb down to $Z=8$. Their individual production cross-sections and recoil velocities at the five energies are presented. Production cross-sections are compared to previously existing data and to empirical parametric formulas, often used in cosmic-ray astrophysics. The experimental data are also extensively compared to different combinations of intra-nuclear cascade and de-excitation models. It is shown that the yields of the lightest isotopes cannot be accounted for by standard evaporation models. The GEMINI model, which includes an asymmetric fission decay mode, gives an overall good agreement with the data. These experimental data can be directly used for the estimation of composition modifications and damages in materials containing iron in spallation sources. They are also useful for improving high precision cosmic-ray measurements.
\\

\vspace{0.5cm}
\noindent

\footnotetext[1]{Present address : IRSN, BP 17, 92262 Fontenay-aux-Roses Cedex, France}
\footnotetext[2]{Present address : GANIL, BP 55027, 14076 Caen Cedex 05, France}
\footnotetext[3]{Present address : CIEMAT, Avda.Complutense,22. 28040 MADRID}
\footnotetext[4]{Present address : National Superconducting Cyclotron
Laboratory, MSU, USA}
\section{Introduction}

The spallation cross-sections of nuclides such as Fe have been historically studied to understand the propagation of cosmic-ray ions in the Galaxy, and to determine the composition of the Galactic Cosmic Ray (GCR) source~\cite{Tsao93}-~\cite{Zeit}.  Galactic cosmic rays constitute a superthermal gas that is partially confined in the Galaxy by interstellar magnetic fields with some leakage into the intergalactic medium.  While propagating in the Galaxy, cosmic rays pass through the interstellar medium and some primary cosmic ray nuclei spallate into secondary cosmic ray nuclei.  As measured by instruments in the solar system, the composition includes both primary cosmic rays whose abundance is depleted by spallation, and secondary cosmic rays produced by spallation.  
As a result of spallation during propagation, certain elements in the GCRs are far more abundant (often by orders of magnitude) than in solar system material.  Examples of these "secondary elements" include Li, Be, B which are mainly spallation products of C and O, and Sc, Ti, V and Cr which are mainly spallation products of Fe.  Conversely, those elements where the abundance of heavier elements is much smaller, and hence have very small secondary contributions are "primary elements."  Prominent examples include C and O and Fe.  Provided the spallation cross-sections are known, the abundance of secondary elements relative to primary elements are a measure of the amount of material cosmic rays traverse in the Galaxy.   This in turn constrains astrophysical models of cosmic rays in the Galaxy.  
It is possible to correct abundance measurements for propagation back to the "source," that is, to determine the composition of the material that became the cosmic rays.  The secondary-to-primary ratios combined with the cross-sections determine the amount of material traversed during propagation in the Galaxy; the amount of material traversed, again with the cross-sections, is then used to correct the measured abundances to the source abundances.  Thus, uncertainties in the cross-sections are more significant than any details of the astrophysical models.  (The exception to this generalization are the unstable secondaries.)  
In recent years, new high resolution elemental and isotopic measurements have become available (i.e. the ACE~\cite{ACE} and Ulysses~\cite{Ulysses} space missions), including measurements in the iron region.  The main source of uncertainties in determining both cosmic-ray secondary production and source composition using these data are uncertainties in the nuclear cross-sections.  The interstellar medium is composed $\sim 90\%$ by number of H atoms and ions.  Most high resolution measurements are of cosmic rays with energies per nucleon in the interstellar medium of $\sim 0.5$ to $\sim 1.5$ GeV.  The cross sections reported here are thus directly applicable to improved interpretation of high-precision cosmic-ray measurements.  

Spallation reactions have also gained a renewed interest with the recent projects of spallation neutron sources and accelerator-driven sub-critical reactors systems considered for the transmutation of nuclear waste (Accelerator Driven Systems (ADS)). In these systems, a high-intensity proton beam of energy around 1 GeV is guided on a spallation target made of a high-mass material. In ADS, neutrons produced in the spallation target are used to maintain the reactivity in the sub-critical reactor where nuclear waste can be transmuted. The proton beam under vacuum in the accelerator has generally to cross a window before entering the spallation target. As it is continuously submitted to the proton beam irradiation, it is one of the most sensitive parts in ADS or spallation-neutron-source design. Among the problems created by the proton irradiation are the changes in the chemical composition of the window material and embrittlement created by gas production and atomic displacements (DPA) in the crystal lattice. A large range of materials have been studied for this window and, in most of the projects, martensitic steels composed at 90 \%  of Iron (with also substantial quantities of Chromium and Molibdenium) have been retained due to their resistance to thermal constraints and radiation effects. Therefore, it is important to have a good knowledge of the production cross-sections of spallation residues in Iron and of their recoil velocity. 

In recent years, an important effort has been undertaken, mainly under the framework of the HINDAS European project~\cite{HINDAS}, to collect a comprehensive set of high-quality spallation data regarding the production of neutrons~\cite{sylvie, NESSI}, light charged particles~\cite{NESSI2} and residual nuclei. The general goal is to better understand the reaction mechanisms in order to improve the models that are then implemented into high-energy transport codes. These codes, validated on experimental data, can afterwards be used to reliably predict all quantities needed for the design of ADS or spallation sources as neutron production, activation or damages.

As concerns residue production, up to now the emphasis was put on spallation reactions on heavy nuclei. Isotopic cross-sections of residues produced in the reactions $^{197}Au + p$~\cite{{Fanny},{Pepe}} at 800 A~MeV , $^{208}Pb + p$ at 1 A~GeV and 500 A~MeV~\cite{{Wla},{Timo},{Bea}, {Laurent}}, $^{238}U + p$ at 1 A~GeV~\cite{{Julian}, {Monique}}, $^{238}U + d$ at 1 A~GeV ~\cite{{Quique},{Jorge}} have already been measured using the reverse-kinematics method at GSI (Darmstadt). In this paper we present new experimental results concerning the isotopic production cross-sections and recoil velocities of spallation residues in the reaction $^{56}Fe + p$ for five energies of the iron beam (0.3, 0.5, 0.75, 1.0 and 1.5 A~GeV). This measurement is the first consistent set of data on isotopically identified residues on a large energy domain and for a light nucleus of practical interest.
The comparison of the obtained data with various models, some of them being quite successful for heavy systems, allows testing  their predicting capabilities for light nuclei and their dependence on  beam energy.

\section{Experimental method}

\subsection{Experimental set-up}

In October 2000, an experiment was performed using the reverse kinematics at GSI in Darmstadt, Germany.
A primary beam of $^{56}Fe$ was delivered by the heavy-ion synchrotron SIS at energies of 0.3, 0.5, 0.75, 1.0 and 1.5
A~GeV and directed onto a liquid-hydrogen target designed and built in the Laboratoire National Saturne (Saclay,
France)~\cite{{Chesny}}.

\bigskip

The liquid-hydrogen thickness was $87.3 mg/cm^2$ contained by titanium windows of $20 \mu m$ each. Two additional Ti foils were used to isolate the vacuum around the target from the vacuum of the beam pipe for security reasons so that a total of $36 mg/cm^2$ of Ti contributes to the empty-target counting. Measurements were repeated with an identical empty target in order to subtract the production on the titanium container from the measured yields of residual nuclei. The contribution of these walls to the counting rates was below 10 \% for
the main part of the residues and below 20 \% for the lightest ones.

The time structure of the primary beam was a pulse of 6 s every 12 s, and the intensity was limited to $10^7$ part/spill. This beam intensity was measured using a secondary-electron emission monitor (SEETRAM)~\cite{{Jur02}} calibrated at the beginning and at the end of each set of measurements at a given beam energy. This was done at low counting rates with a plastic scintillator as absolute reference.

\bigskip

\begin{figure}[h!tb]
\begin{center}
\includegraphics[width=16.0cm]{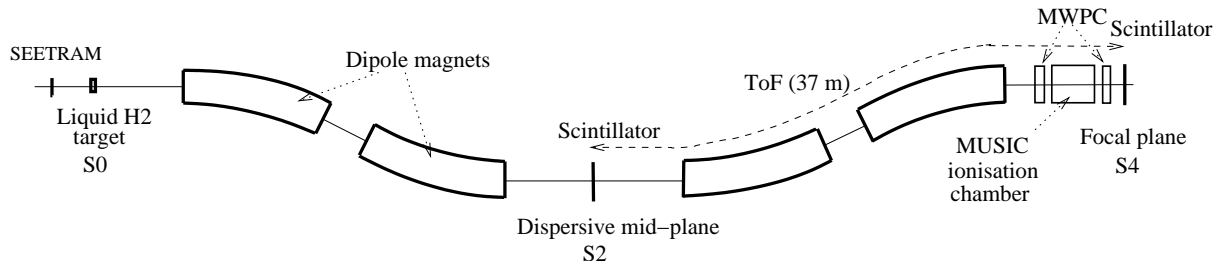}
\end{center}
\caption{{Schematic layout of the FRS fragment spectrometer. Fragments are
analyzed by the four large dipole magnets. Scintillators at S2 and S4 measure the time of flight over
the second half of the spectrometer as well as the horizontal positions in the dispersive focal planes at S2 and at S4. The MUSIC detector (ionization chamber) gives information about the energy loss of the fragment. Multi-Wire Proportional Chambers (MWPC) are used for beam tuning and removed for production measurements.}}
\label{fig:frs}
\end{figure}

Residual nuclei produced in the reaction with the target were focused in the beam direction and analyzed with the
FRS (Fragment Separator)~\cite{{FRS}} operated as an achromatic magnetic spectrometer. {Fig.}~\ref{fig:frs} is a schematic diagram of the experimental setup showing the four large dipole magnets and the essential detector equipment.

\bigskip

\bigskip

Due to their  relativistic energies, the fragments produced in this experiment are fully stripped. The horizontal positions of these ions and a time of flight (ToF) were measured with two plastic scintillators, one located in the intermediate dispersive plane S2 and the other one installed at the final achromatic focal plane S4. The signal from the scintillator at S4 was used as the trigger for the acquisition of all detectors. The nuclear charge Z was determined using a multiple-sampling ionization chamber (MUSIC). The energy loss in the gas produces a signal proportional to $\frac{Z^2}{\beta^2}$, allowing the determination of Z with a resolution of $\Delta Z=0.3$ (FWHM) charge units.

\bigskip

The knowledge of the horizontal positions of the ions determines precisely the radii $\rho 1$ and $\rho 2$ of their trajectories in the two magnetic sections of the spectrometer. An absolute calibration is obtained with the iron beam detected in specific measurements at low intensity. Together with the magnetic field strengths in the dipoles measured with Hall-effect probes, the  magnetic rigidities $B \rho 1$ and $B \rho 2$ can be determined for each ion. Therefore, a total identification of the nature of the ions could be performed from the relation :

\bigskip

\begin{equation}
\frac{A}{Z}=  \frac{e B \rho}{m_u c \beta \gamma}
\end{equation}

\bigskip

where $m_u$ was the atomic mass unit and $\beta$ $\gamma$ were deduced from the experimental time of flight.
Note that in this formula we have replaced the mass of the $(A,Z)$ ion by $A.m_u$ which means neglecting binding energies compare to nucleon masses. 

\bigskip

\begin{figure}[h!tb]
\begin{center}
\includegraphics[width=11.0cm,height=10.0cm]{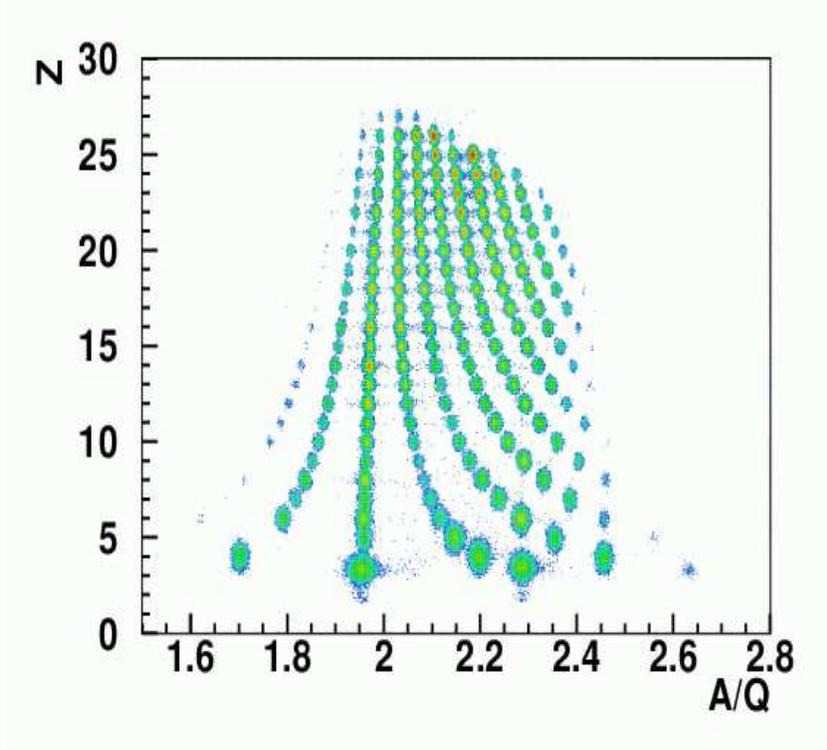}
\end{center}
\caption{{Complete isotope coverage in Z vs. A/Q (actually identical to A/Z) for 1 A~GeV $^{56}Fe$ on the liquid-hydrogen target. The plot
is built from data of overlapping settings, normalized to the primary beam intensity and superimposed.}}
\label{fig:nucleos}
\end{figure}

The FRagment Separator has a momentum acceptance of $\pm 1.5$\%. Therefore, about 18-20 different settings of the FRS were needed to cover the complete velocity distribution of all the ions. {Figure}~\ref{fig:nucleos} shows the complete fragment coverage in the Z vs. A/Z plane for 1 GeV per nucleon $^{56}Fe$ on the hydrogen target. The plot was made by adding histograms from individual settings, each one normalized to the dose of the primary beam. Fragments are well resolved and easily identifiable in this plot down to lithium. However, for the lightest elements the transmission of the spectrometer is very low, necessiting a dedicated method of analysis. This has been done only at 1 GeV per nucleon and reported in a separate paper ~\cite{{Paolo}}. Therefore, we show in this paper results of the production cross section and recoil velocity only down to Z=8-10, depending on the beam energy considered.  
\bigskip

\subsection{Data analysis}

The fragments are first identified in Z using the ionization chamber, taking into account the position and velocity dependence of the energy-loss signal. The velocity distribution of the fragments is obtained with high precision using the time-of-flight and magnetic-rigidity measurements. The experimental time-of-flight between the intermediate and the final focal plane is precise enough for an unambiguous identification of the fragment mass. After identification of the isotope, a more accurate value of the longitudinal velocity can be deduced from the magnetic rigidity in the first part of the spectrometer using relation 1.

\bigskip

\begin{figure}[h!tb]
\begin{center}
\includegraphics[width=9.0cm,height=9.0cm]{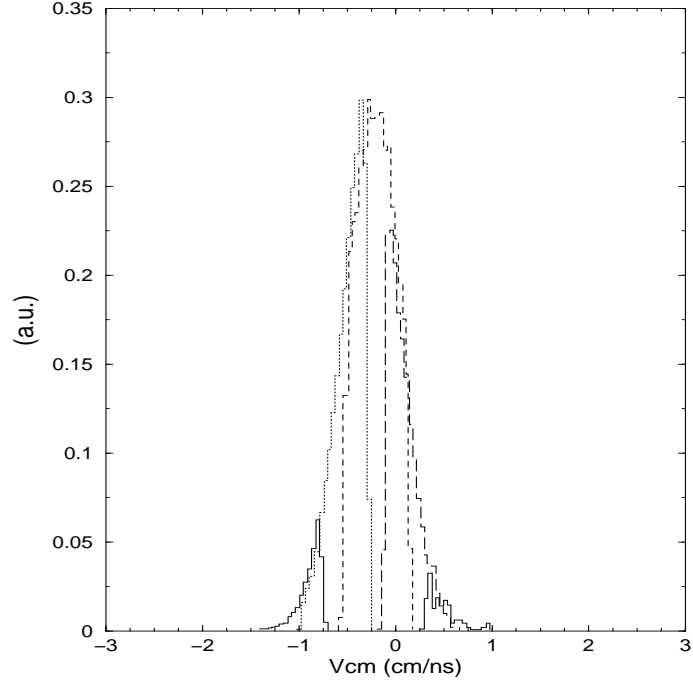}
\end{center}
\caption{{Longitudinal velocity distribution of $^{38}K$ detected as a residual nucleus at 1 A GeV and expressed in the rest frame of the iron beam. Five different
settings of the FRS were needed to reconstruct the complete distribution. The yield (here in arbitrary units) detected in each setting is
normalised by the number of incident iron nuclei and corrected for the acquisition dead time.}}
\label{fig:potasio}
\end{figure}

Assuming that the reaction takes place at the center of the target, the fragment velocity is corrected for the energy
loss in the target and transformed into the reference frame of the projectile at rest. A measurement of the recoil velocity of the fragments is thus obtained in that frame. To obtain the production cross-section of a given isotope, it is necessary to reconstruct the full velocity distribution by adding the partial ones measured in different settings, with the proper normalization. An example of the velocity distribution for $^{38}K$ is shown in {Fig.}~\ref{fig:potasio}. For this isotope, five different settings of the FRS were needed in order to reconstruct the complete velocity distribution.

\bigskip

\bigskip

Due to potential damages in the detectors, isotopes having a magnetic rigidity too close to the beam one could not be measured. This
is why the detection of $^{54}Mn$ was not possible in this experiment. For the same reason, some settings of other isotopes could not be obtained, leading to truncated measured velocity distributions. In that case a fit by a Gaussian function excluding the truncated zones was used to reconstruct the full distribution and then determine the total cross-section, the mean value of the velocity and its variance. In the case of a truncated zone in the velocity distribution too large to have a converging fit, the parameters of the Gaussian were constrained using the neighboring isotopes. The reconstruction procedure leads to an uncertainty on both the velocity determination and the isotope production cross-section. These uncertainties have been estimated by taking into account the fluctuation of the reaction point in the target and by doing reasonable variations of the fitted parameters for several groups of isotopes.

\subsection{Corrections and uncertainties}

The isotopic production cross-section of each spallation residue $\sigma(Z,A)$ was obtained from the difference between the yield measured with the hydrogen target $(Y_H (Z,A))$ and the yield measured with the empty target $(Y_e(Z,A))$, each of them corrected for their dead time (correction factor $f_{\tau H}(f_{\tau e})$) and normalized to the number of incident iron nuclei $N_{Fe\ H}(N_{Fe\ e})$.

\bigskip

\begin{equation}
\sigma(Z,A)= \left( \frac{Y_H(Z,A) \cdot f_{\tau H}}{N_{Fe\ H}} - \frac{Y_e(Z,A) \cdot f_{\tau e}}{N_{Fe\ e}} \right) \cdot
\frac{f_{\epsilon} \cdot f_{trans} \cdot f_{sec}}{N_H}
\end{equation}

The cross section is finally obtained after a division by the number of hydrogen nuclei per surface unit $N_H$ and with additional corrections due to the detection efficiency ($f_{\epsilon}$), the transmission of the FRS ($f_{trans}$) and the secondary reactions ($f_{sec}$) estimated for hydrogen events. It was determined that, even at the lowest energy, a correction for possible change of charge state is not necessary.

\bigskip

Losses of events due to the dead time of the experiment, mainly due to the acquisition capability, are estimated for each run from the ratio between the free triggers measured on a scaler of high counting-rate capability and the number of recorded events (or accepted triggers). During the experiment, the counting-rate conditions were kept so that this correction never exceeded 30\%, and was most frequently smaller for detection at magnetic rigidities substantially different from the beam rigidity.

\bigskip

An estimation of the global detection efficiency $f_{\epsilon}$ including the detailed analysis of all needed information can be obtained from the difference between the number of accepted triggers and the final number of events that have been analyzed. An event can be analyzed if all the elements required have been registered without any problem: position at the two focal planes, time of flight and energy loss in the
MUSIC detector. The trigger signal obtained by a narrow coincidence on high signals produced by highly ionizing particles is here supposed to identify a true heavy ion with a probability of nearly 100\%. In almost all settings this efficiency was in the range 96-99\%.

\bigskip

Corrections due to secondary reactions in the target and in the layers of matter on the trajectory of the fragments (mainly the plastic scintillator of 3 mm thickness at S2) were calculated following the method described in~\cite{{pagkh}} as previously used in other similar experiments~\cite{{Fanny}}. If a second reaction occurs in the target, the initially produced ion becomes lighter, so that cross sections of light ions are artificially increased (and the one for the corresponding heavy ion decreased). If a reaction occurs in the plastic at S2, the spallation ion will most often be out of the narrow magnetic rigidity acceptance in the second part of FRS and so will be lost at S4. Total nuclear interaction cross sections for the different fragments were estimated using the parametric formula of Kox et al~\cite{{Kox}}. The maximum value (8\%) of this correction factor is obtained for the secondary reactions in the target leading to the lightest evaporation residues. It decrease to zero for heavy residues. The correction due to the lost in the scintillator if a reaction occures is of the order of 3.5\% and was taken into account (as a function of the nature of the ion and of it's mean energy). The attenuation of the beam flux inside the finite target thickness was also taken into account in this correction and is equal to -2\% for a reaction cross section of 700 mb.

\bigskip

The transmission correction is the most important factor concerning losses in the detection. Due to its geometrical characteristics and the ion optics, the FRS has only an angular acceptance of $15$ mrad around the beam axis, and a large number of the fragments analyzed in this experiment have an angular distribution at the entrance of the FRS larger than this acceptance. An evaluation of the fraction of the residual yield not detected in the experiment had to be made from the measured velocity distribution of the fragment as it is described in~\cite{{trans}}.
Considering that, in the projectile reference frame, the emission of the fragments can be described as a 3-D Gaussian distribution around a mean longitudinal recoil, the width of the angular distribution in the laboratory frame can be obtained from the longitudinal velocity distribution measured in the experiment:

\bigskip
\begin{equation}
\sigma(\theta)\approx \frac {\sigma(v_{\parallel})}{<v_{\parallel}>}
\end{equation}

{\noindent where $<v_{\parallel}>$ is the mean value and $\sigma(v_{\parallel})$ the width of this distribution for evaporation residues of a given mass.}

\bigskip

The transmission through the FRS can be parameterized as :

\bigskip

\begin{equation}
T=1-exp(- \frac{\alpha_{eff}(x2,x4)^2}{2 \sigma(\theta)^2})
\end{equation}

{\noindent where $\alpha_{eff}(x2,x4)$ is the effective angular acceptance of the FRS as a function of the ion positions $x2$  and $x4$ respectively at the intermediate S2 and the final focal planes S4. This angle was calculated with the code described in reference~\cite{{trans}} using 15 mrad as the maximum angular acceptance when the ion optics is the most appropriate.}

\bigskip

\begin{figure}[h!tb]
\begin{center}
\vspace{1cm}
\includegraphics[width=8.0cm,height=7.0cm]{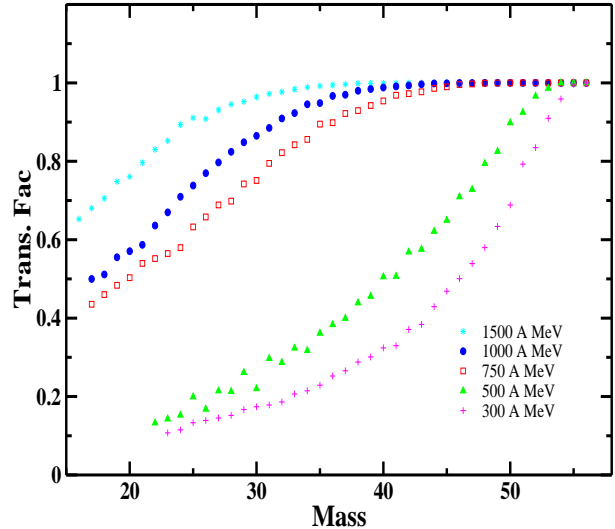}
\end{center}
\caption{Transmission factor as a function of the mass number of the residue for the five energies presented in this work (see text).}
\label{fig:transmision}
\end{figure}

The transmission factor varies from 1 (no correction) to 0.4 for the lightest fragments that have a much larger angular distribution (see {Fig.}~\ref{fig:transmision}) for the three highest energies. Various reasonable assumptions on the calculation of $\alpha_{eff}(x2,x4)$ lead to uncertainty estimations on $T$ of $1\%$ to  $15\%$ for the lightest evaporation residues. However, the analysis has revealed that at 500 and 300 MeV/A, the magnetic optics settings used during the experiment was not optimal and that the maximum acceptance of the FRS was reduced to $9.15$ mrad. This value has been taken into account in the transmission factor leading to much larger corrections for these two energies as it can be seen in {Fig.}~\ref{fig:transmision}.

\bigskip

For the absolute normalization, the precision on the target thickness has been studied in previous experiments ~\cite{{mustapha}} and is estimated to be $2.5\%$. The absolute numbers of incident ions $N_{Fe\ H}$ and $N_{Fe\ e}$ for runs with the hydrogen target and the empty target respectively are obtained from the SEETRAM calibration with an absolute error estimated to be $2.8\%$.

 Experimental values for the isotopic cross-sections with their errors are listed in appendix A. The $^{54}Mn$ that could not be measured was obtained by a smooth interpolation between the neighboring isotopes so the value given in the tables is followed by (Interp.). This value is used to obtain integrated quantities as the mass or charge distributions and in the evaluation of the total reaction cross section also given in appendix A.

 \bigskip

Final results of the mean recoil velocity and the width of the velocity distributions for the various residual nuclei are presented in appendix B. Errors quoted here are due to the velocity reconstruction procedure above described and to the magnetic-rigidity determination. In the case of a truncated velocity distribution, results partially interpolated are followed by (I). The minus sign means that the recoil velocity is opposite to the original direction of the iron beam or in other words in the direction of the proton motion in the iron at rest system.

\section{Results}

\subsection{Isotope production cross sections}

Using the experimental method described above it was possible to measure at five different energies most of the residues produced in the spallation reaction of iron with cross-sections larger than $10^{-2}$ mb, from cobalt (Z=27) down to oxygen (Z=8) or neon (Z=10) depending on the energy. At 1 A~GeV, cobalt isotopes have not been measured.

\bigskip

\begin{figure}[h!tb]
\begin{center}
\includegraphics[width=10.5cm,height=9.5cm]{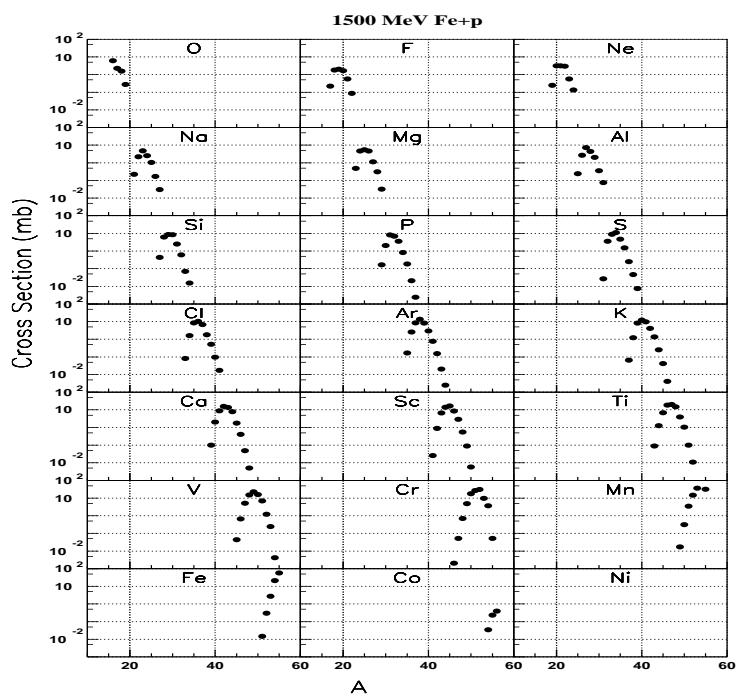}
\end{center}
\caption{{Isotopic production cross-sections of fragments from the reaction $^{56}Fe +p $ at 1.5 A~GeV as a function of mass number}}
\label{fig:iso1500}
\end{figure}

\begin{figure}[h!tb]
\begin{center}
\includegraphics[width=10.5cm,height=9.5cm]{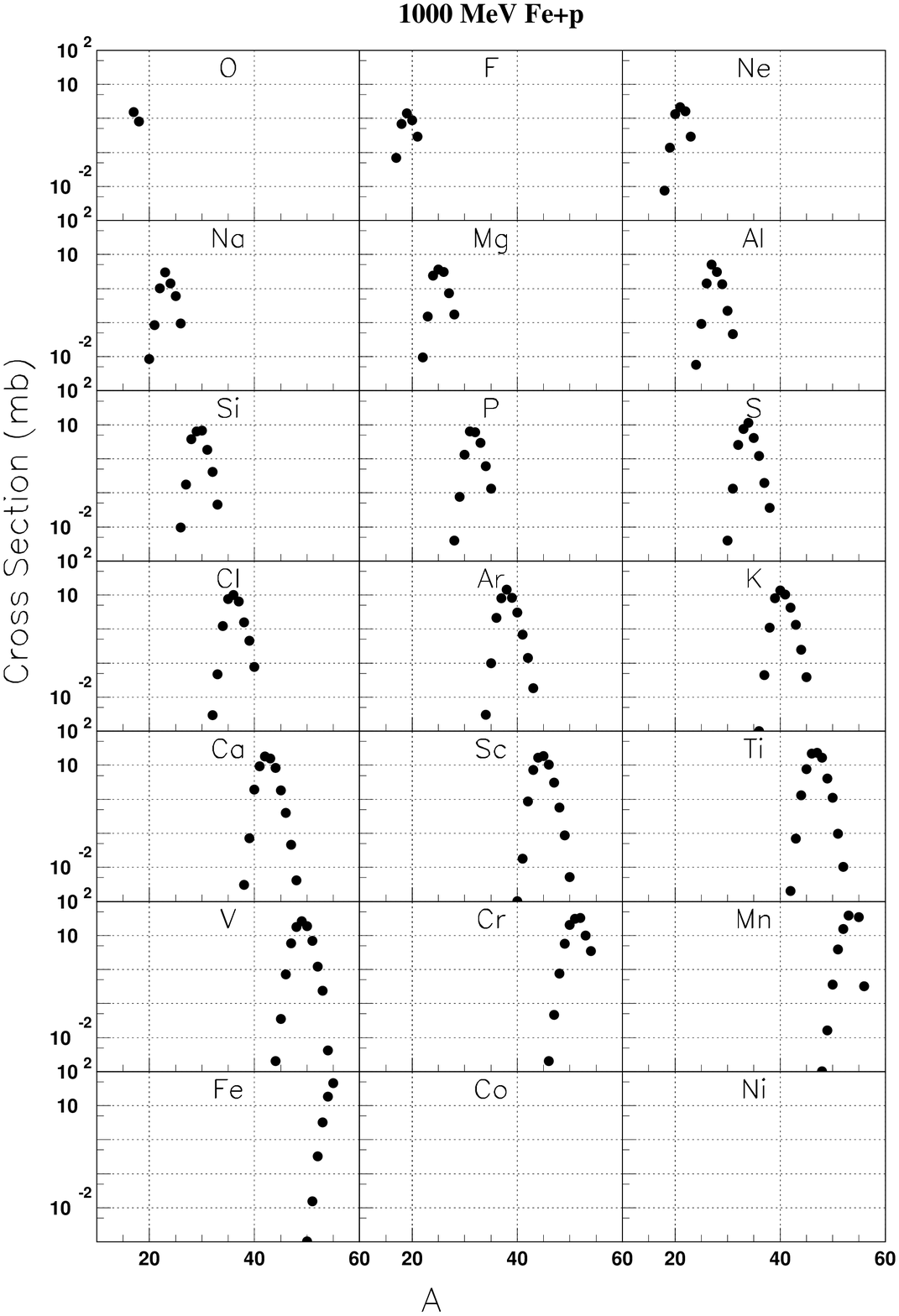}
\end{center}
\caption{{Isotopic production cross-sections of fragments from the reaction $^{56}Fe +p $ at 1.0 A~GeV as a function of mass number}}
\label{fig:iso1000}
\end{figure}

\begin{figure}[h!tb]
\begin{center}
\includegraphics[width=10.5cm,height=9.5cm]{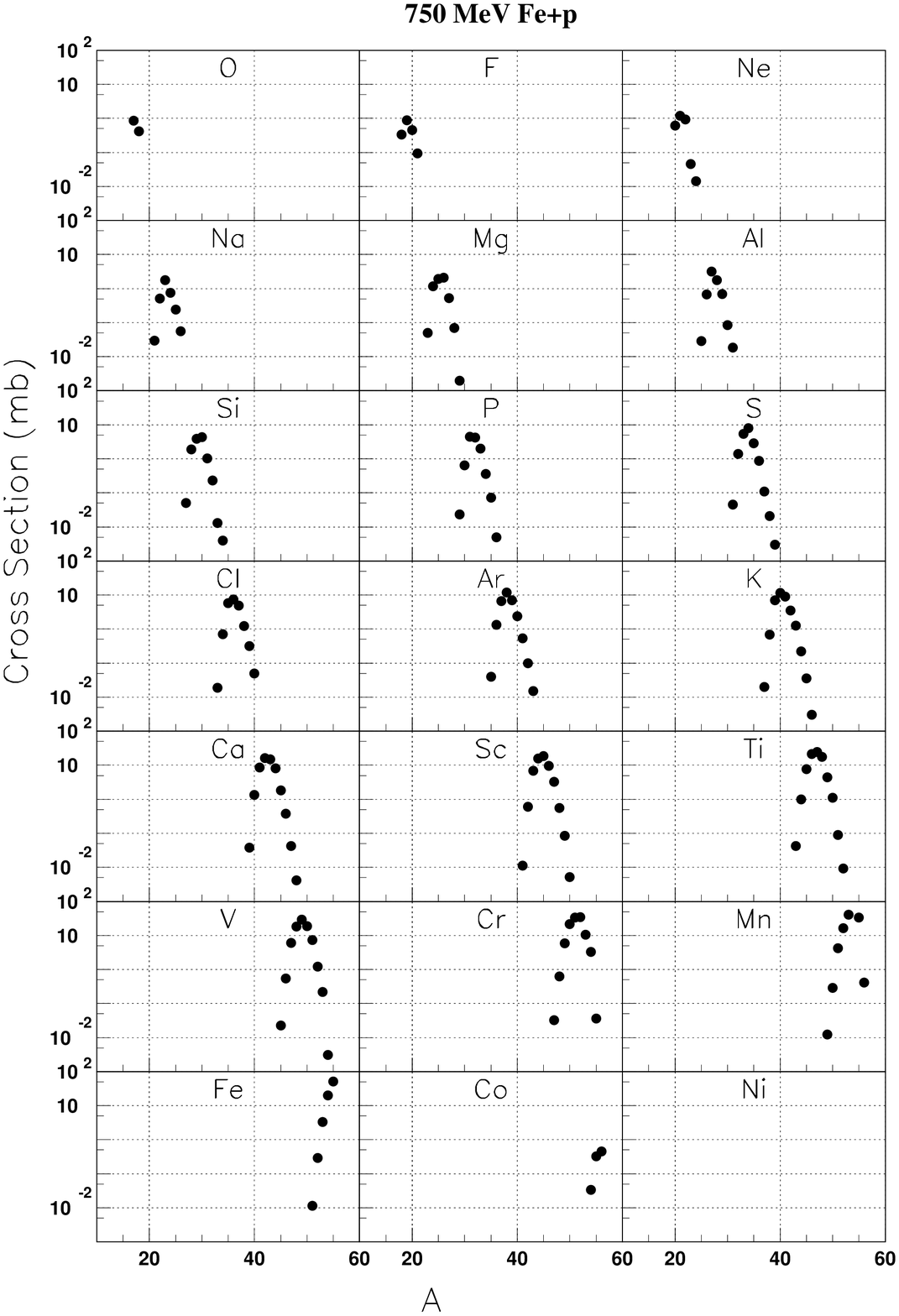}
\end{center}
\caption{{Isotopic production cross-sections of fragments from the reaction $^{56}Fe +p $ at 0.75 A~GeV as a function of mass number}}
\label{fig:iso750}
\end{figure}

\begin{figure}[h!tb]
\begin{center}
\includegraphics[width=10.5cm,height=9.5cm]{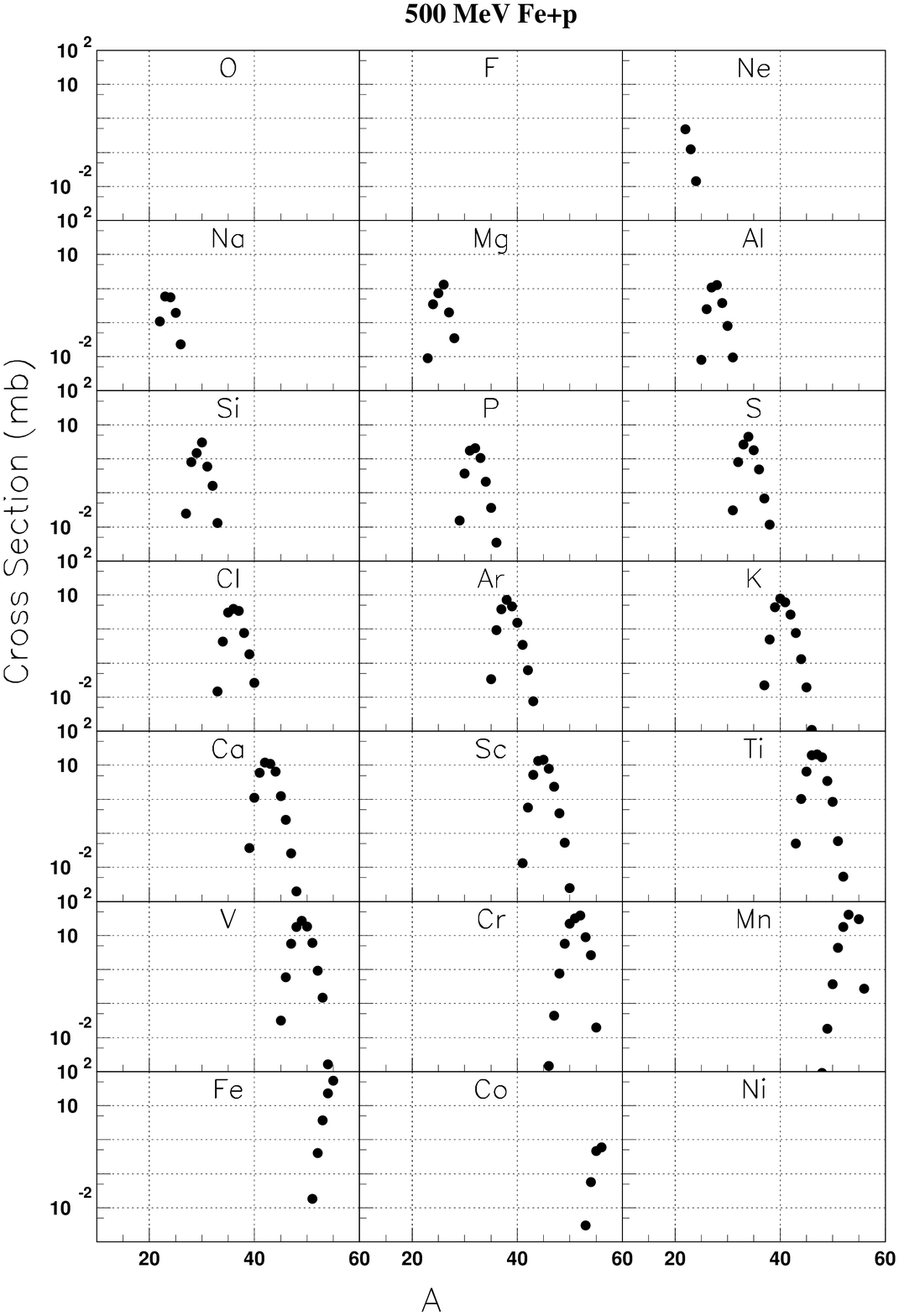}
\end{center}
\caption{{Isotopic production cross-sections of fragments from the reaction $^{56}Fe +p $ at 0.5 A~GeV as a function of mass number}}
\label{fig:iso500}
\end{figure}

\begin{figure}[h!tb]
\begin{center}
\includegraphics[width=10.5cm,height=9.5cm]{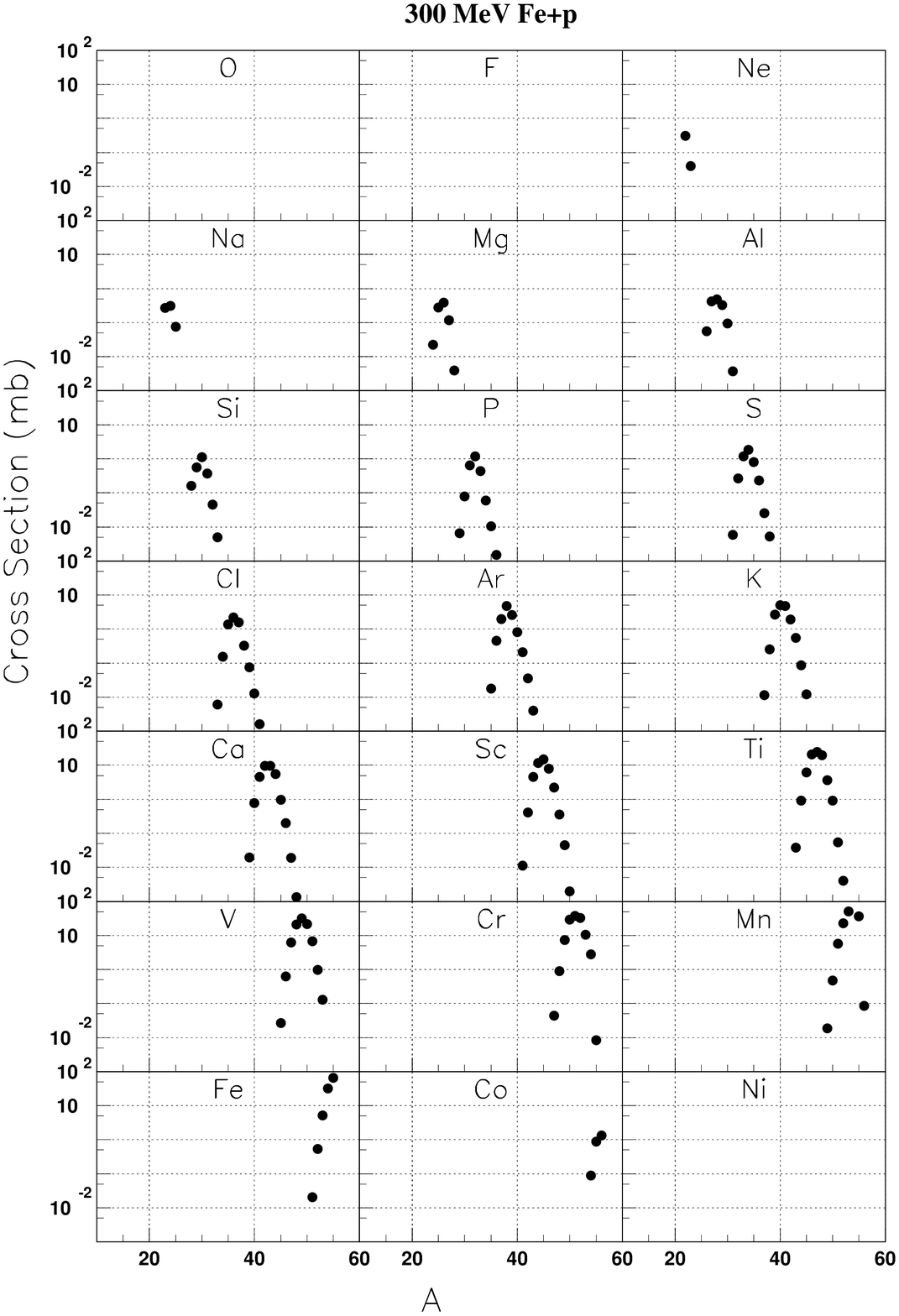}
\end{center}
\caption{{Isotopic production cross-sections of fragments from the reaction $^{56}Fe +p $ at 0.3 A~GeV as a function of mass number}}
\label{fig:iso300}
\end{figure}

{Figures}~\ref{fig:iso1500}, \ref{fig:iso1000}, \ref{fig:iso750}, \ref{fig:iso500} and \ref{fig:iso300} show the isotopic distribution cross sections at the five beam energies. Error bars do not appear as they are smaller than the data points. The position of the maximum of these isotopic curves is correlated with the excitation energy transferred in the collision between the projectile and the target. In the case of a peripheral collision, in which the excitation energy is limited, only a few particles are evaporated by the fragment, leading to the population of isotopes close to stability. For more central collisions, the deposited excitation energy is larger and more neutron-deficient isotopes are produced due to the evaporation phase which favors the emission of neutrons. However, the tendency towards neutron-deficient isotopes is weaker than what is generally observed in heavy systems since, for iron, the Coulomb barrier is much smaller and the neutron to proton ratio in the projectile is also smaller.

\bigskip

\begin{figure}[h!tb]
\begin{center}
\includegraphics[height=9.5cm]{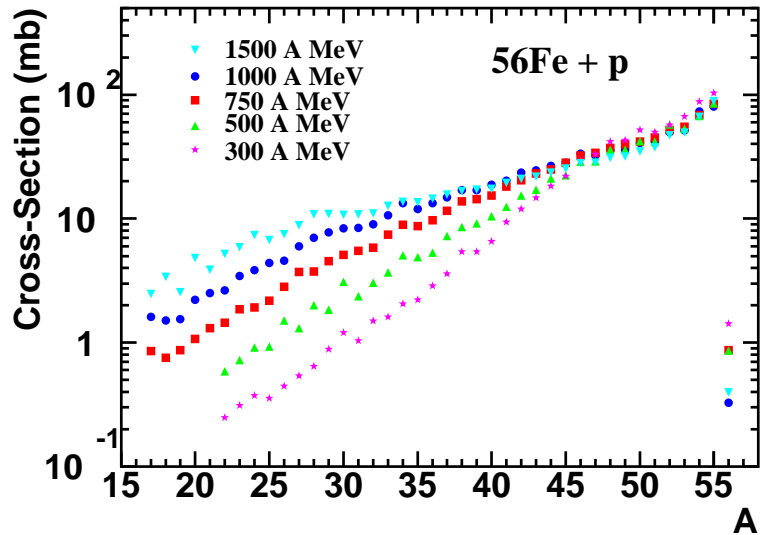}
\end{center}
\caption{{Mass distribution of the residual nuclei in the spallation reaction $^{56}Fe + p$ at the five different beam energies.}}
\label{fig:distria}
\end{figure}

Isotopic cross-sections can be summed to obtain mass or charge distributions. {Figure}~\ref{fig:distria} presents the mass distribution of the spallation residues for the five energies of the iron beam analyzed in this experiment. The residues are produced with different cross-sections depending on the energy of the projectile. The general trend of the data is globally as expected. As the beam energy increases, the deposited excitation energy becomes more and more important, leading in average to a stronger evaporation of nucleons, and finally to lighter evaporation residues. This is reflected by the substantial rise of the light fragment cross-sections between 300 and 1500 MeV per nucleon. As the total reaction cross section is overall rather constant over the studied energy range, this is compensated by a decrease of the production cross sections of the heaviest evaporation residues with increasing energy. It appears that masses around 46-47 are produced with a cross section almost independent of the beam energy.

\subsection{Comparison with other experimental data}

\subsubsection{Reverse kinematics}

The present data can be compared with the ones obtained by W. R. Webber and collaborators using the reverse kinematics method. Measurements were
performed on either a thick CH2 target (from $\sim 2 g/cm^2$ to $\sim 6 g/cm^2$) subtracting the carbon
contribution~\cite{{Web90},{Webform},{Web902}}, or a liquid-hydrogen target ($1.52 g/cm^2$)~\cite{{Web},{Web1086}} at SATURNE. In both cases, the fragments were detected with a telescope of scintillators and Cerenkov counters.

\begin{figure}[h!tb]
\begin{center}
\includegraphics[width=9.0cm,height=9.0cm]{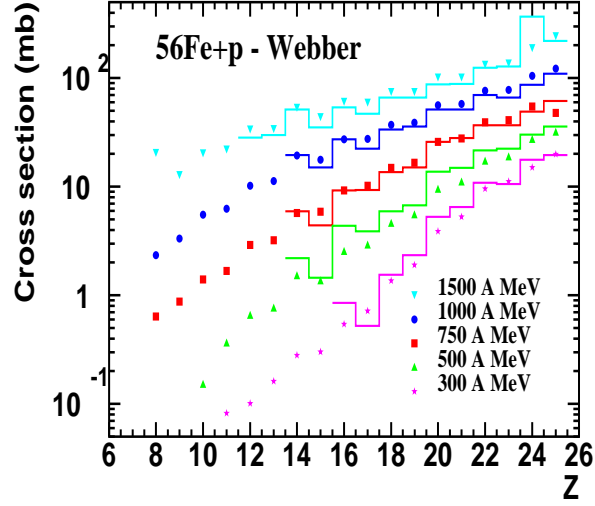}
\end{center}
\caption{{Nuclear-charge distribution of the residual nuclei for the five energies with scaling factors (2/1/0.5/0.25/0.125 respectively from 1500 A~MeV to 300 A~MeV) applied for clarity. Points correspond to the present data, and solid histograms are data from Webber et al. ~\cite{{Web902},{Webform},{Web902}} at close energies : 1512, 1086, 724, 520 and 330 A~MeV.}}
\label{fig:webz}
\end{figure}

The charge distributions of the spallation residues for several iron beam energies from 330 to 1615 A~MeV ~\cite{{Web1086},{Webform},{Web902}} have been measured down to Z around 15. In {Fig.}~\ref{fig:webz}, these results (histograms), at beam energies close to ours, namely 1512, 1086, 724, 520 and 330 A~MeV are compared with the present cross-sections (symbols) summed over masses to obtain the charge distribution.
The overall agreement is satisfying in terms of variation with energy and charge of the residue. A systematic dependence of the element cross sections with the parity of Z is consistently observed in both experiments. The deviation factor, i.e. the average ratio between the two experiments has been calculated and is shown in {Table}~\ref{tab:webfac}. The cross-section for Z=24 at 1512 A~MeV, for which the Webber value is much larger than the neighboring cross sections and inconsistent with a general trend, is excluded. At the three highest energies, it is perfectly compatible with the precisions of both experiments (5\% to 20\% for Webber et al. and 9\% to 15\% here). At 300 A~MeV (330 A~MeV), the discrepancy is larger but still acceptable considering the different energies (10\%) of the two measurements. The highest value (1.28) for the deviation factor is found at 500 A~MeV (520 A~MeV). Although this could be caused by a particular experimental problem at this energy, it is still compatible within the respective errors, especially if one bears in mind that at low energy both errors are larger: in our case because of the large transmission correction and in the case of Webber because of corrections for secondary reactions. The same reasons could explain the fact that, for a given energy, the disagreement is increasing with decreasing Z values, as it can be seen in {Fig.}~\ref{fig:webz}. Another argument is that if we plot charge-changing cross sections as a function of the beam energy for various charges, our results at 500 A~MeV are $\sim $10\% below a smooth interpolation based on the other measured energies whereas the Webber values are $\sim $20\% above the interpolation.

\begin{table}[h]
\vspace{0.5cm}
\small
\centering
\begin{tabular}{|l|c|c|c|c|c|}
\hline
Energy/A (MeV) & 300  & 500  & 750  & 1000  & 1500  \\
\hline
Deviation factor & $1.23 $ & $1.28 $ & $1.01 $ & $0.89 $ & $0.88 $ \\
\hline
\end{tabular}
\caption{\label{tab:webfac}\textit{Average ratio of the charge-changing cross-sections measured by Webber et al.~\cite{Web1086, Webform, Web902} divided by the values from this experiment.}}
\vspace{1.0cm}
\end{table}

\bigskip

\begin{figure}[h!tb]
\begin{center}
\includegraphics[width=9.0cm,height=9.0cm]{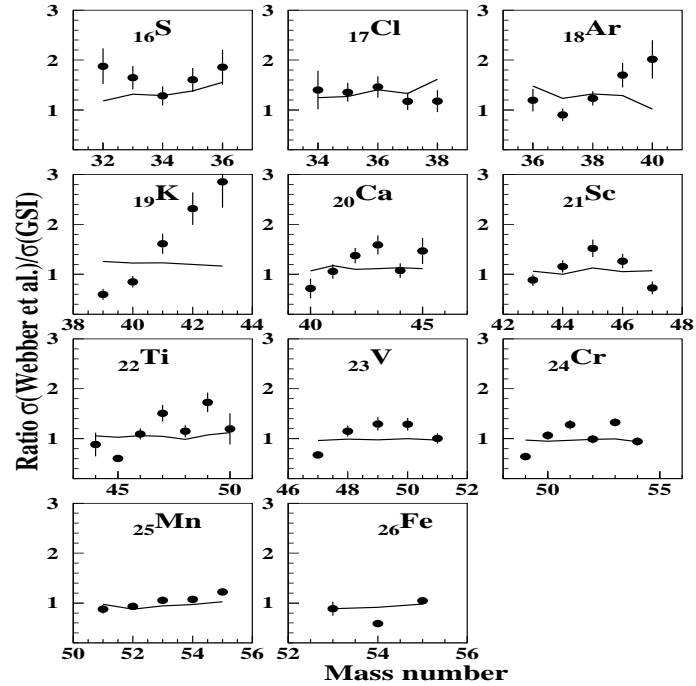}
\end{center}
\caption{{Ratio between isotopic cross-sections measured by Webber et al. at 573~A~MeV~\cite{{Web},{Web1086}} and the present experiment at 500 A~MeV for each element as a function of the mass number. Lines are theoretical predictions from INCL4-GEMINI for this ratio.}}
\label{fig:ratiowebber}
\end{figure}

The isotopic production cross-sections have also been measured previously but only at one energy (573 A~MeV), using a liquid-hydrogen target~\cite{{Web},{Web1086}} and were limited to rather large cross-sections. The ratio between these values and the present data is displayed in {Fig.}~\ref{fig:ratiowebber}, including the respective errors. The lines represent the ratios of the cross sections at 573 A~MeV and 500 A~MeV computed with the INCL4-GEMINI combination of models. It shows that the difference in energy between the two experiments is not negligible for the lightest fragments, for which it can lead to differences of 30 to 40\%. The agreement between the data is quite good for residues close in mass to iron but the difference increases for lighter isotopes. The value of the ratio is frequently hardly compatible with the expected value given by the line.

\bigskip

\begin{figure}[h!tb]
\begin{center}
\includegraphics[width=9.0cm,height=5.0cm]{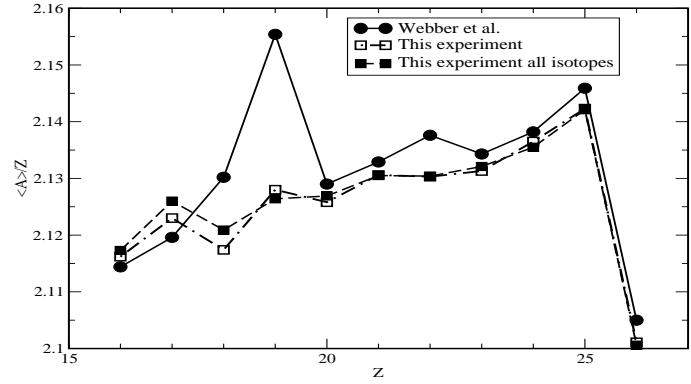}
\end{center}
\caption{{Ratio of the mean nuclear-mass value to the charge for each element measured by Webber et al. at 573~A~MeV~\cite{{Web},{Web1086}} (black circles) and in the present experiment at 500 A~MeV (open squares). Black squares are for the same quantity evaluated from all isotopes measured in this experiment.}}
\label{fig:webber_Amean.Z}
\end{figure}

Actually, one would expect a smooth variation of the mean value and of the width of the isotopic distribution with element charge. In {Fig.}~\ref{fig:webber_Amean.Z} are represented the mean mass-over-charge ratio as the function of Z, summing only the isotopes measured by both experiments. Clearly these quantities are more fluctuating in the Webber et al. experiment, in particular for potassium (Z=19) data and to a smaller extent for argon (Z=18) and titanium (Z=22) ones. The use of our full isotopic distributions, which extend much beyond the ones of Webber et al., does not make a large difference.

\bigskip

\subsubsection{Direct kinematics}

\begin{center}
\begin{figure}[h!c]
\begin{center}
\includegraphics[width=14.0cm,height=17.0cm]{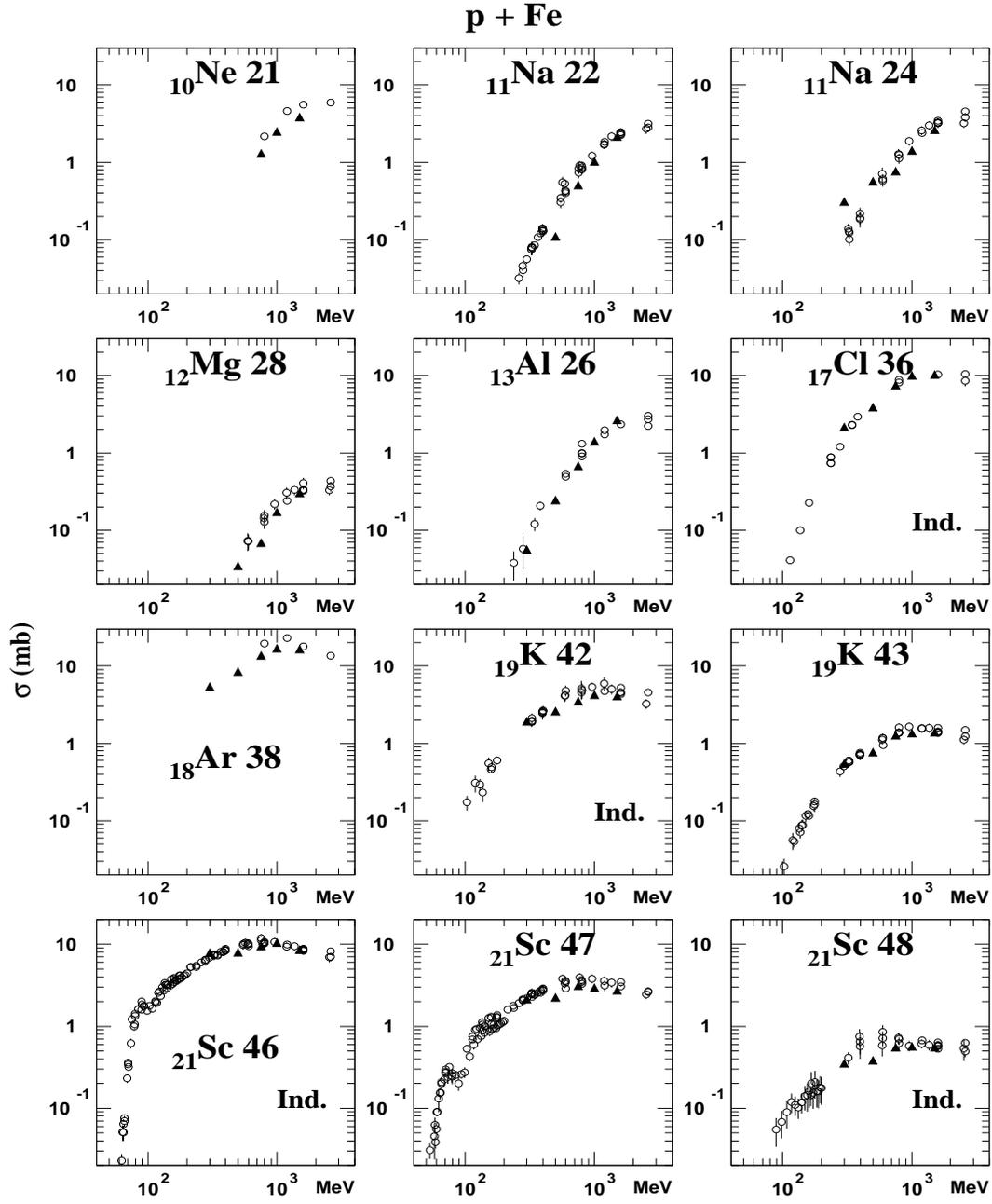}
\end{center}
\caption{{Excitation functions of some residual nuclei produced in the spallation reaction of proton on iron. Open dots are the data of R. Michel et al. obtained by a direct irradiation~\cite{{Mich97},{Mich04}} and black triangles
correspond to the present experimental data at 5 energies. Independent isotopes are indicated (Ind.).}}
\label{fig:fe1}
\end{figure}
\end{center}

\begin{center}
\begin{figure}[h!c]
\begin{center}
\includegraphics[width=14.0cm,height=17.0cm]{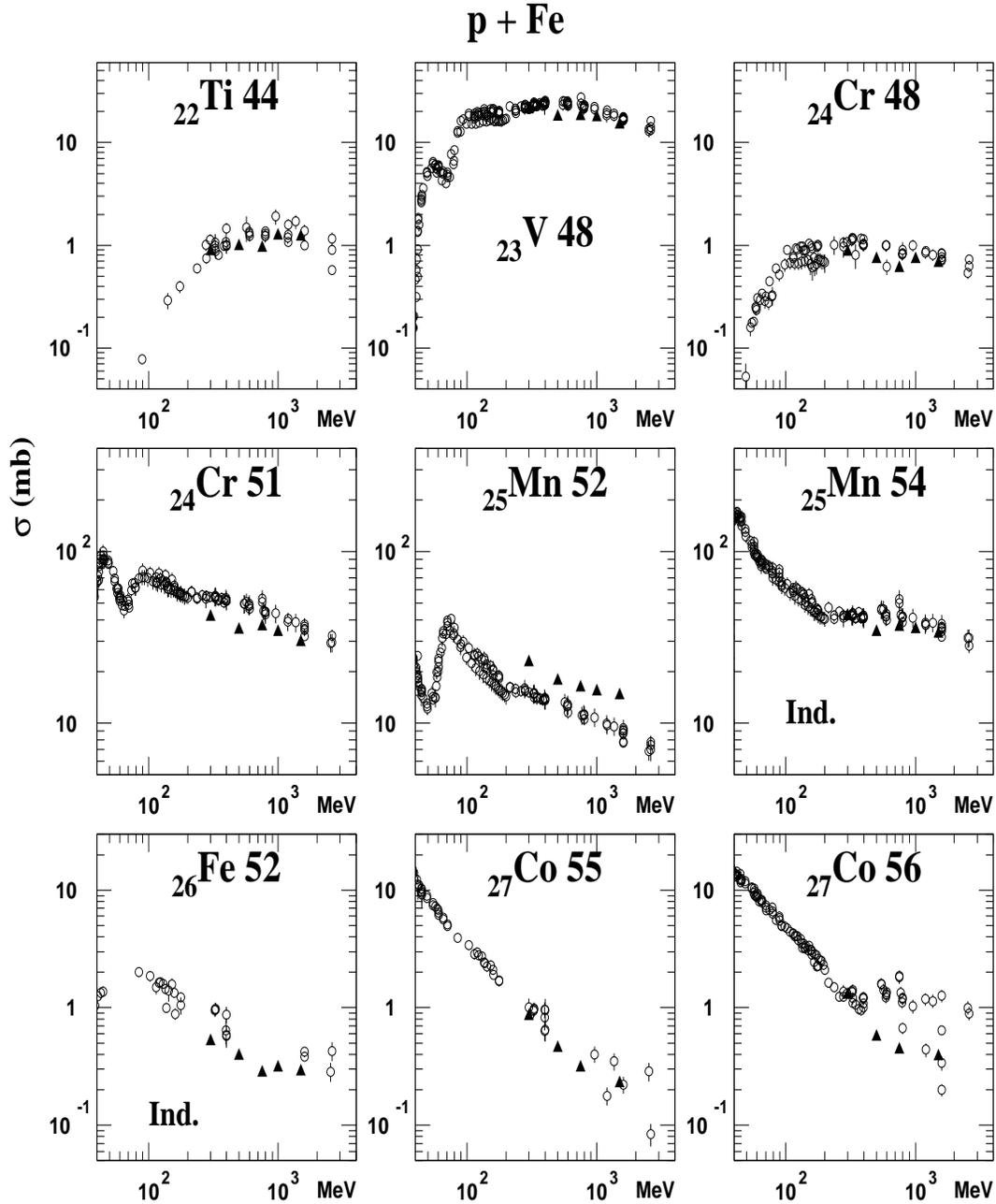}
\end{center}
\caption{{Excitation functions of some residual nuclei produced in the spallation reaction of proton on iron. Open dots are the data of R. Michel et al. obtained by a direct irradiation~\cite{{Mich97},{Mich04}} and black triangles
correspond to the present experimental data at 5 energies. Independent isotopes are indicated (Ind.).}}
\label{fig:fe2}
\end{figure}
\end{center}

Results in direct kinematics have been obtained by R. Michel and collaborators~\cite{{Mic},{Mich97},{Mich04}} by irradiation of natural iron targets at different proton beam energies, allowing the determination of excitation functions from a few tens of MeV to about 2 GeV. Some of the produced residual nuclei have been measured and identified by their gamma-ray decay spectrum or by mass-spectrometry. These data are compared to our experimental data in {Fig.}~\ref{fig:fe1} and {Fig.}~\ref{fig:fe2}. Results can be split into  ``cumulative'' and ``independent'' nuclei meaning that they are or not populated by a decay chain. In the case of cumulative cross-sections, our own cross-sections have been summed along the decay chain before comparing with Michel's data. The following isotopes: $^{36}Cl$ $^{42}K$, $^{46}Sc$, $^{48}Sc$ $^{54}Mn$ and $^{52}Fe$ are ``independent''.


Our data follow quite well, in most of the cases, the dependence on energy obtained in the Michel et al. experiment. This is very satisfying if we consider the difference between the two experimental methods. Some of the important differences that can be noticed may be due to the use of natural iron in the case of Michel's data. For instance, the observed higher cross-section for $^{52}Fe$ could come from a contribution of the (p,2n) reaction on $^{54}Fe$ adding to the (p,4n) on $^{56}Fe$. Although there is only 6\% of $^{54}Fe$ in natural iron, the effect should be non-negligible since (p,2n) is 40 times more probable than (p,4n) as deduced from our results. Conversely, the lower cross-sections found by Michel for $^{52}Mn$ and the higher one for $^{56}Co$ at high energy do not seem compatible with the tendency deduced from our isotopic distributions.

\clearpage

Finally, we can say that the present results are qualitatively in good agreement with former measurements. The fact that we have complete isotopic distributions extending down to lighter nuclei than previously measured, on a wide range of energy, allows us to check the consistency of our own results and detect possible inconsistencies in other sets of data.

\subsection{Comparison with parametric formulas}

Since 1950, parametric formulas have been developed by astrophysicists with the aim of predicting the production cross-sections of the residual nuclei in spallation reactions. These formulas are used in case of light and intermediate nuclei present in the composition of the cosmic-rays like iron. In this section we present the comparison of our new experimental data with the results of three of these parametric formulas:
Webber~\cite{Webber}, EPAX~\cite{Summerer} and Silberberg and Tsao~\cite{{ST90},{ST98}} formulas.

\bigskip

\subsubsection{Webber's formula}

This parametric formula has been developed by Webber et al.~\cite{Webber} from the experimental data shown in the previous section. It is
used in case of light spallation residues with $Z_i<28$ and for energies of the projectile $E>200$ MeV.

\bigskip

The form of this formula is :

$$
\sigma(A_i,Z_i,E) = \sigma_0(Z_i, Z_t) \cdot f_1(Z_i,A_i, Z_t, A_t) \cdot f_2(E,Z_i,Z_t)
$$

for residual nuclei $(Z_i,A_i)$ of the spallation reaction on a target nuclei $(Z_t,A_t)$ at energy $E$.

\begin{itemize}
\item The first factor $\sigma_0(Z_i,Z_t)$ gives the charge distribution of the residues
\item $f_1(Z_i,A_i, Z_t, A_t)$ describes the isotopic curves (from their data at 573 MeV per nucleon)
\item $f_2(E,Z_i,Z_t)$ gives the energy dependence
 \end{itemize}
\bigskip

\begin{figure}[h!tb]
\begin{center}
\includegraphics[width=10.0cm,height=8.0cm]{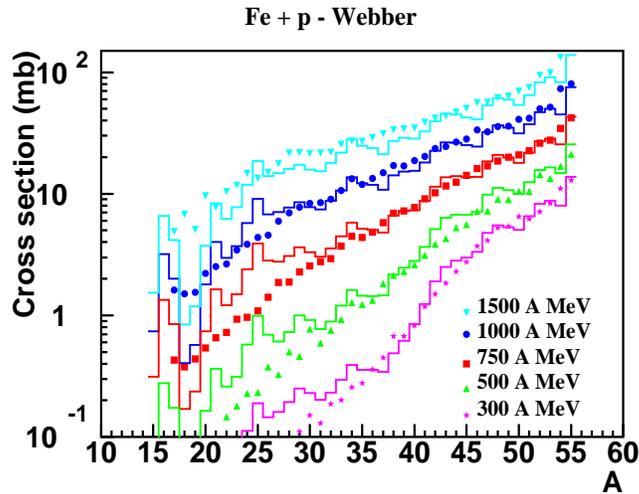}
\end{center}
\caption{{Comparison between present results at 5 energies (symbols) and the results obtained with Webber's formula (solid lines). Scaling factors (2/1/0.5/0.25/0.125 respectively from 1500 A~MeV to 300 A~MeV) are applied for clarity.}}
\label{fig:paramweb}
\end{figure}

\bigskip

In {Fig.}~\ref{fig:paramweb} the comparison of our mass distribution with the predictions of the Webber's formula is shown. A rather good agreement is obtained at all the energies considered here for the heaviest residues, which are precisely those already measured by Webber et al. and used to determine the parameters of the formula. However, there seems to be some oscillations in the cross-sections not observed in the data. Actually, the charge distribution, not shown, is more accurately predicted by the formula. This comes from the fact that the isotopic distributions predicted by the Webber's formula have smaller widths than those obtained experimentally (see {Fig.}~\ref{fig:paramshape}). A probable explanation is that only very few isotopic data were existing at the time when the formula was established. Therefore, the isotopic dependence could not be properly determined. Furthermore, the extrapolation of the parametric formula for light residues that are measured here for the first time shows an important discrepancy with the data. Even if this parametric formula can be useful for determining the production of the most produced spallation residues, this illustrates the danger of using parametric formulas outside the range in which they were adjusted.

\subsubsection{The EPAX formula}

Epax formula~\cite{Summerer} was created with the aim of describing the production of residues in fragmentation reactions between heavy ions in what is call the \textit{limiting-fragmentation regime} in which the residue production cross-section does not depend anymore on the energy of the projectile. Although not fully valid for protons at these energies, it might be instructive to know how close its predictions are to the present data. The limiting-fragmentation regime for the spallation reaction Fe + p is expected to be reached for energies of a few GeV per nucleon so here we can just expect the 1.5 A GeV data to be comparable with it.

It can be used for spallation reactions with protons in the case of target nuclei of masses $18< A_t < 187$, although developed mainly for heavy-ion reaction . It is composed by two factors :

$$
\sigma(Z_i, A_i)= Y_A \cdot \sigma(Z_{prob}-Z_i)
$$

with :

\begin{itemize}
\item $Y_A$ a factor to describe the mass distribution of the fragments
$(Z_i, A_i)$\\

$$
Y_A= S_2 (A_t^{1/3} + A_{pro}^{1/3}+S_1) \cdot P \cdot  exp[-P (A_t-A_i)]
$$

and $lnP=P_2 \cdot A_t +P1$. $S_1$, $S_2$, $P_1$ and $P_2$ being adjusted
parameters and $A_{pro}$ the mass of the projectile (one here for protons).

\item $\sigma(Z_{prob}-Z_i)$  describes the isobaric curves with $Z_{prob}$ as the charge for the maximal production. The various
$Z_{prob}$ values as a function of A defines the so called \textit{residue corridor} in this approach.
  \end{itemize}

\begin{figure}[h!tb]
\begin{center}
\includegraphics[width=10.0cm,height=8.0cm]{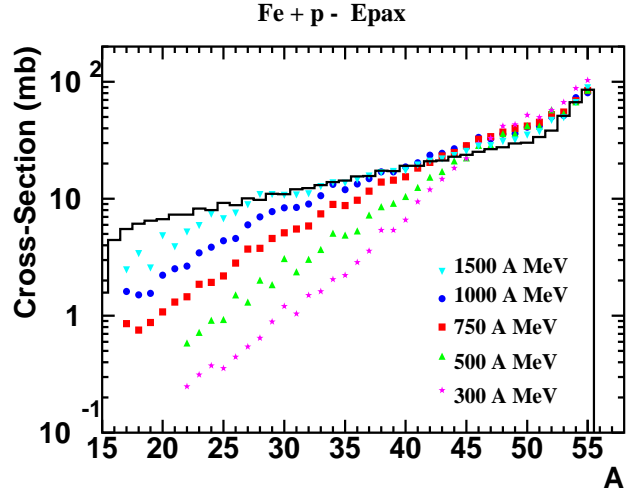}
\end{center}
\caption{{Comparison between the present measured mass distributions (symbols) and the
results obtained with the EPAX formula (solid line). }}
\label{fig:paramepax}
\end{figure}

In {Fig.}~\ref{fig:paramepax} our experimental results (symbols) are compared with the predictions of the EPAX formula. The experimental data at 1.5 GeV per nucleon are expected to be the ones closest to the limiting fragmentation regime, therefore we have renormalized the factor $S_2$ so that the formula gives the total reaction cross-section measured at 1.5 GeV per nucleon (794 mb). Since the EPAX total cross section was 617 mb, this led to a multiplication by 1.28.

It can be seen, as expected, that as the energy increases the mass distribution gets closer and closer to the EPAX prediction, with a quite good agreement at 1.5 GeV per nucleon. However, the lightest residues are still overestimated by the formula. The EPAX formula predicts also a more important evaporation of neutrons than seen in the isotopic cross-section data. In fact the measured N/Z ratio of the residues is higher than the one of the residue corridor which is used in the formula.

\subsubsection{Silberberg and Tsao's formula }

\bigskip

The first version of this parametric formula  has been developed  in 1973~\cite{ST73} with the experimental data measured by Rudstam~\cite{Rud} concerning the spallation residues in the spallation reaction $p + Fe$ at 340 MeV. Various improvements, especially the beam-energy dependence, have been added in successive versions~\cite{{ST90},{ST98}}. It can be written as :

\bigskip

$$
\sigma(A,Z,E) = \sigma_0 \cdot f(A) \cdot f(E) \cdot e^{-P(E) \Delta A} \cdot e^{-R | Z-SA-TA^2 | ^ {\nu}} \Omega \cdot \eta \cdot  \Phi
$$

\bigskip

where :

\bigskip

\begin{itemize}
\item $\sigma_0$ is a normalization to the total reaction cross-section
\item $f(A)$ and $f(E)$ are factors used only in the case of target nuclei $Z_t > 30$
\item $e^{-P(E) \Delta A}$ represents the reduction in the production cross-section with the mass difference ($\Delta A$) between the
residue and the target nuclei and an energy dependence through the $P$ parameter
\item $ e^{-R | Z-SA-TA^2 | ^ {\nu}}$ describes the width and the position of the maximum in the isotopic and isobaric production
\item $\Omega$ takes into account the level structure of the residual nuclei
\item $\eta$ is a factor for the pairing of protons and neutrons
\item $\Phi$ represents an increase in the production of very light residues
\end{itemize}

\begin{figure}[h!tb]
\begin{center}
\includegraphics[width=10.0cm,height=8.0cm]{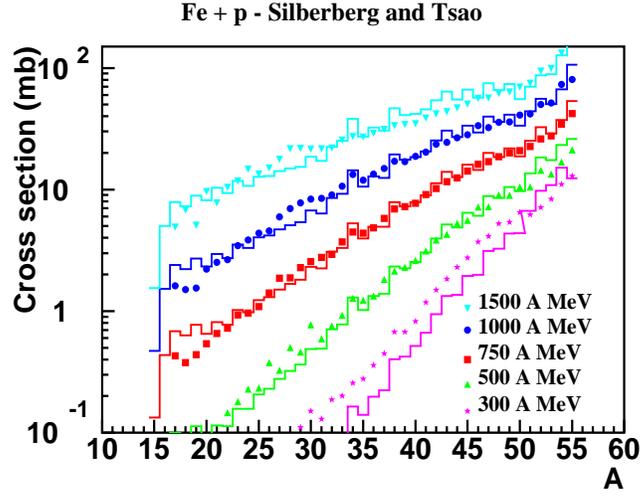}
\end{center}
\caption{{Comparison between present experimental results at the five energies (symbols) and the
results obtained with the Silberberg and Tsao's formula (solid lines). Scaling factors (2/1/0.5/0.25/0.125 respectively from 1500 A~MeV to 300 A~MeV) are applied for clarity. }}
\label{fig:paramst}
\end{figure}

In {Fig.}~\ref{fig:paramst} a comparison of this formula with the experimental results presented in this work is shown. In general, the agreement is very good for all energies between 10 \% and 30 \% at 300 MeV per nucleon where the discrepancy is larger.

This last parametric formula appears as the most suitable to reproduce the present data, probably because of the largest data base used to derive it, which contains systems rather close to the ones studied here. These formulas are quite useful for quickly calculating production rates. Although some physical ingredients are present to derive them, more sophisticated approaches are needed to better handle the physics included in spallation reactions and to describe more fully other observables than cross-sections.

\subsubsection{Isotopic distribution shapes}

\bigskip

\begin{figure}[h!bt]
\begin{center}
\includegraphics[height=9cm]{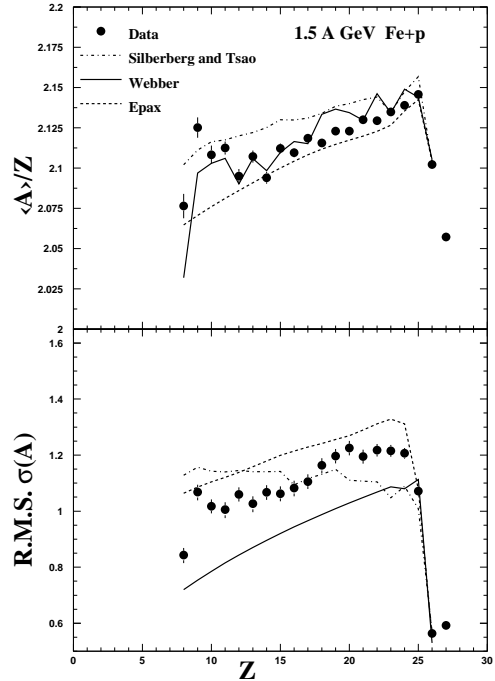}
\end{center}
\caption{{Mean values and width of the mass-over-charge distributions as a function of element charge measured at 1500 MeV per nucleon compared with the predictions of the parametric formulas of Webber (solid line), Silberberg-Tsao (dashed-dotted line) and EPAX (dashed line).}}
\label{fig:paramshape}
\end{figure}

In the preceding sections only mass distributions were compared to the predictions of the parametric formula. It is also interesting to know how they reproduce the isotopic distributions. A powerful way to look at this is to compare the shape of the mass distributions of each element through the mean value and width of the mass-over-charge distributions as a function of Z. This is what is shown in {Fig.}~\ref{fig:paramshape} in which the experimental results at 1500 MeV per nucleon (for better chance of agreement of EPAX) are compared with the three parametric formula. It can be seen that, as concerns the mean mass-over-charge, Webber's formula and EPAX agree rather well with the data while Silberberg-Tsao's predicts a slightly too high value. Regarding the widths, EPAX is acceptable and Webber tends to produce a too narrow mass distribution, maybe because the formula was fitted on his isotopic data which have a rather limited extension. Silberberg-Tsao gives a nearly constant width with Z, in contradiction to the experimental shape. This means that this formula, which gave the best agreement for mass distribution, should be used with caution if one wants to estimate isotope production cross-sections.

\subsection{Comparison with models}

The design and optimization of spallation sources requires the knowledge of a large number of quantities directly related to spallation reactions in different materials and at various energies. Since exhaustive measurements of such a large amount of data are beyond the experimental possibilities, one needs to develop spallation models with a good predictability that can be used in transport codes for simulations. This implies a deeper knowledge of the physics of the spallation reactions.

\bigskip

Spallation is generally described by a two-step mechanism. The first stage of intra-nuclear cascade process (INC) governed by nucleon-nucleon collisions, leads to an excited nucleus after the ejection of a few energetic particles (p, n, $\pi$, d, $\alpha$ etc.). The second longest phase follows corresponding to the evaporative decay of the excited remnant nucleus with a possible competition with fission and Fermi break-up in some cases. Some approaches include also an intermediate stage of pre-equilibrium to account smoothly for the transition to the full thermalization of the evaporating nucleus.

Old INC models are still currently used in the high-energy transport codes employed for applications as Bertini~\cite{Bertini} or ISABEL~\cite{Isabel}) models. However, recently, a renewed interest for INC codes has been triggered by new available spallation data. Among them one could cite recent improvements on the INC codes found in~\cite{INCL4, mashnik}. In the present work, we have compared the experimental results of the spallation residues on iron to the predictions of three INC codes: Bertini code, ISABEL and INCL4~\cite{INCL4}. Since a long time, the first two ones are available in transport codes like LAHET3~\cite{lahet} and MCNPX~\cite{mcnpx} for simulations of macro-systems, Bertini (with pre-equilibrium) being used by default. INCL4 was only recently implemented in these code systems as well as CEM~\cite{mashnik}. The basic physical assumptions are rather similar but differ in their implementation, for instance the way to develop the NN series of interactions, the way to treat Pauli blocking or the criterium to stop the INC stage. Note that we have used the implementation of ISABEL in LAHET3 which is blocked above 1 GeV. But this does not means that this cascade is not valid at higher energies.

For the second stage of the reaction, the most commonly used de-excitation code (and default option) in LAHET and MCNPX is the Dresner evaporation code~\cite{Dresner} complemented with the Atchison model for fission~\cite{Atch}. It uses the Weisskopf-Ewing formalism~\cite{Weisskopf} for the treatment of the evaporation, as do the more recent models ABLA~\cite{ABLA} and GEM~\cite{GEM}. Mainly, these three codes differ in the formulas and parameters used to described the level densities, the Coulomb barriers and the inverse reaction cross-sections. The Dresner code includes only the evaporation of light particles: neutrons, hydrogen and helium isotopes. The ABLA code has been mainly tuned for heavy systems with a particular interest on the fission description. In the version used in this work only neutrons, protons and alpha particles are evaporated. Furthermore shell and pairing effects as well as gamma decay were not taken here into account. The GEM code is a recent update of the Dresner model with new parameters and extends the Weisskopf-Ewing formalism to the evaporation of intermediate-mass fragments up to Z = 12. Actually the three codes (Dresner, ABLA and GEM) do not take into account in the evaporation process the angular momenta, which in fact are relatively small in spallation reactions induced by incident protons. Fission of heavy systems is described in a Bohr and Wheeler approach using phenomenological fragment distributions in Atchison and GEM. In ABLA fission is treated as a dynamical process taking into account the nuclear viscosity, and the fragment distribution is essentially obtained through the calculated population of states above the mass-asymmetric conditional saddle point.

As will be shown in the following, conventional Weisskopf-Ewing evaporation may be not sufficient to account for our data. This is why we will also compare our results to models predicting the emission of intermediate-mass fragments through other mechanisms. The GEMINI model~\cite{gemini} treats evaporation of light particles within the Hauser-Feshbach formalism~\cite{hf}, taking explicitly into account the angular momentum. Following the idea of Moretto~\cite{moretto} that there should be a continuous transition between evaporation and fission, for all systems including light ones, the emission of intermediate fragments is handled as asymmetric fission in the Transition State Model. The transition between Hauser-Feshbach evaporation and asymetric fission can be chosen through a parameter: in the present work, this parameter has been set so that the Transition State Model is used for fragments above helium. Several other options exist in the code. We have used the ones recommended by the author. Some tries to vary them, although not exhaustive, do not reveal strong differences in the description of the present data.

The SMM code is a numerical implementation of the Statistical Multifragmentation Model from~\cite{smm} often used to describe heavy-ion collisions in which multifragmentation is more likely to arise. The parameters to describe the multifragmentation process are the standard ones as described in~\cite{eos}. In particular, the asymptotic freeze-out volume is three times the initial one. The evaporation is treated in the Weisskopf-Ewing formalism up to fragment mass 18, and the lightest primary fragments decays are treated by the Fermi break-up~\cite{fermi}.

In the comparison between experimental data and model predictions, it is always difficult to disentangle the respective roles of the intra-nuclear cascade, which determines the characteristics of the remnant nucleus (charge, mass, angular momentum and excitation energy) at the end of the cascade stage, and of the de-excitation model. For instance, the under-prediction by the INCL4-ABLA combination of models of the light evaporation residue cross-sections observed for heavy systems~\cite{Wla, INCL4} could be ascribed either to a too low excitation energy given by INCL4 or to a deficiency of ABLA at the highest excitation energies. However, some observables can be found that are more sensitive to one reaction stage or the other. In the following, we will try, as far as possible, to disentangle the influences of the intra-nuclear cascade and of the de-excitation stage in the comparison with the different observables.

\subsubsection{Total reaction cross-section}
\begin{figure}[h!tb]
\begin{center}
\includegraphics[width=9.0cm,height=7.0cm]{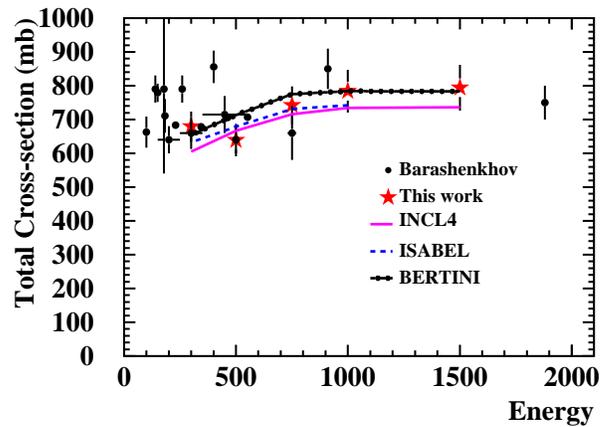}
\end{center}
\caption{{Total reaction cross-sections of protons on iron as a function of the bombarding energy. Our five experimental
data are compared to the compilation of previous experimental data from Barashenkhov~\cite{secreac} and the values given by the three INC codes : Bertini, ISABEL (not available
for E $>$ 1 GeV in LAHET3) and INCL4.}}
\label{fig:setotal}
\end{figure}

The total reaction cross-section is clearly one of the observables that depends only on the INC model since it is mainly related to the probability that the incident nucleon makes a collision with one nucleon of the target and that this collision is not blocked by the Pauli principle. In {Fig.}~\ref{fig:setotal} we present the total reaction cross-sections obtained for the five energies analyzed in this work. They were calculated by summing up the isotope productions tabulated in the appendix.  The summation has been done down to Z=8-10 depending on the bombarding energies. The contribution of the unmeasured isotopes have been estimated to be at most a few percents, i.e. smaller than the error bars. The fact that the lightest fragments could come from binary breakups and therefore leads to a possible double counting in the total reaction cross-section is also negligible. Actually, the two contributions play in opposit directions and even more or less compensate. Previous experimental data from the Barashenkhov compilation~\cite{secreac} are also shown on this figure. A reasonable agreement is observed between most of the previous data and the present ones for both the absolute values and the behavior with the incident energy. The predictions of all the three INC codes agree with the data within the experimental accuracy, the difference between them being at most 10 \%. This is not surprising since these INC models are known to generally well reproduce the total reaction cross-sections at energy above a hundred MeV~\cite{INCL4, lahet}. This observable cannot be used to discriminate between these three codes.

\bigskip
 \subsubsection{Mass and charge distributions}

In this section we examine the various model predictions compared to the mass or charge distributions obtained by summing the measured isotopic cross-sections. For completeness the light fragment cross sections analyzed in~\cite{Paolo} and obtained during the same experiment are also included at 1 GeV per nucleon.

\begin{figure}[h!tb]
\begin{minipage}[c]{17.4cm}
\begin{center}
\includegraphics[width=10cm]{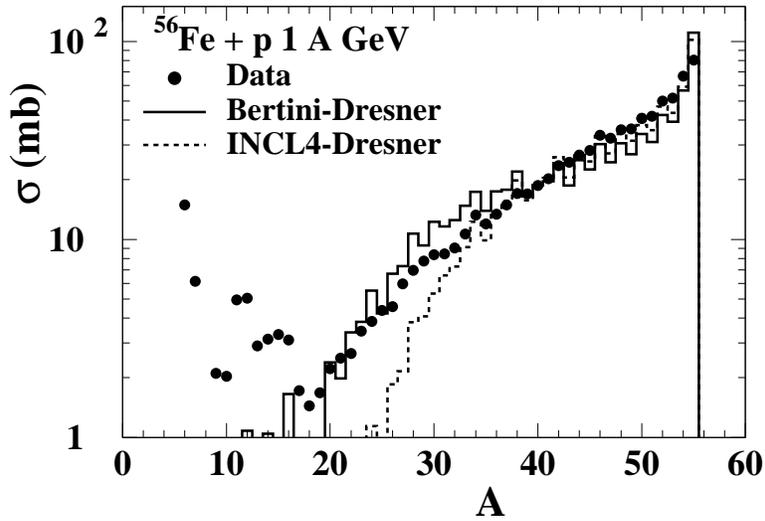}
\end{center}
\caption{{Mass distribution of the spallation residues of iron at 1 A~GeV compared to the Bertini and INCL4 INC models combined with the Dresner evaporation code.}}
\label{fig:bertdres}
\end{minipage}
\end{figure}

\begin{figure}[h!tb]
\begin{center}
\includegraphics[width=10.0cm,height=8.0cm]{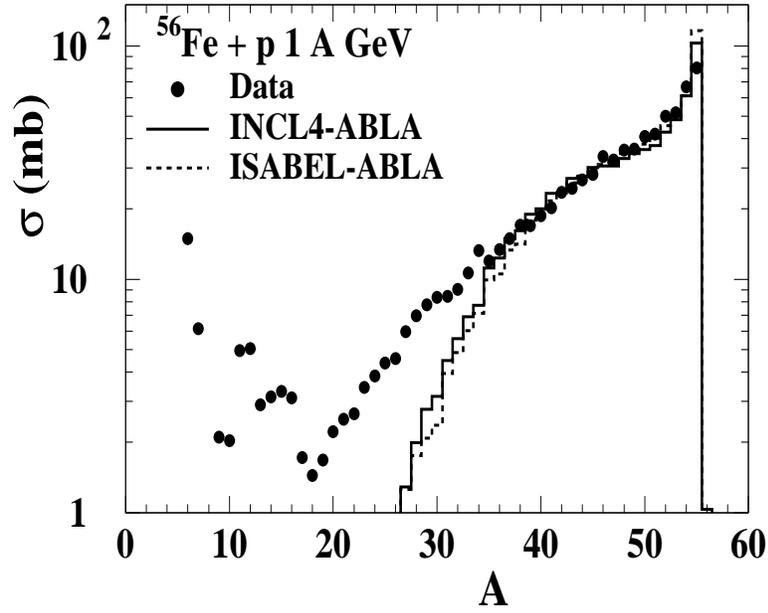}
\end{center}
\caption{{Mass distribution of the spallation residues of iron at 1 A~GeV compared to two different INC codes
(INCL4~\cite{{INCL4}}
and ISABEL~\cite{{Isabel}}) combined with the
ABLA evaporation code ~\cite{{ABLA}}.}}
\label{fig:compcasc}
\end{figure}

We first investigate the influence of the choice of the INC model. In {Fig.}~\ref{fig:bertdres} the mass and charge distributions of the residual nuclei produced at 1 GeV is shown and compared to the Bertini intra-nuclear cascade (plus pre-equilibrium) followed by the Dresner evaporation. Both mass and charge distributions lead to the same conclusions. The production yields of residues close to iron which are the major part of the spallation cross-section are underestimated while the yield of intermediate-mass residues is on the contrary overpredicted. The same conclusions were already obtained for heavy nuclei~\cite{Wla}. This behavior could be ascribed to a too high excitation energies at the end of the Bertini intranuclear cascade even after the introduction of a preequilibrium phase. A comparison is also shown with INCL4 followed by the same evaporation code. The calculations now predicts less excited remnants and a more satisfactory agreement is obtained for the heaviest residues but the light ones are still underestimated. It can be also noticed that, in both cases, the production of very light fragments is by far underpredicted. Another comparison is shown in {Fig.}~\ref{fig:compcasc} between the mass distribution of the spallation residues and the predictions of two different INC models, now ISABEL and INCL4, followed by the ABLA evaporation. This last combination has been shown to reproduce satisfactorily many spallation data~\cite{INCL4} in a wide domain and without adjustment of parameters. Both calculations give a similar good descriptions of the residues close to iron and an underprediction of intermediate and light nuclei cross-sections. This underprediction of INCL4-ABLA is in fact consistent with light evaporation residue cross-sections obtained from heavier nuclei (lead and gold) ~\cite{INCL4}. Actually, for the heaviest nuclei which are mainly formed in peripheral collisions with low excitation energy, the evaporation plays a less important role than the intra-nuclear cascade since only a very little number of nucleons is evaporated. The fact that both calculations have the same behavior and are rather good for heavy residues suggests that the underprediction of the light residues is not due to a lack of excitation energy. Indeed, we have seen in the comparison with Bertini in {Fig.}~\ref{fig:bertdres} that a larger excitation energy does lead to a larger production of light fragments but to the detriment of heavy ones which cannot be counterbalanced by playing with evaporation models. This rather indicates that the problem comes from the de-excitation stage. In the following we will not consider anymore the Bertini model for which many shotcomings have been pointed out~\cite{sylvie},~\cite{Wla},~\cite{ab}. We will mainly restrict the comparisons with various de-excitation models using INCL4 in the first stage, since ISABEL generally gives similar results.

\bigskip

\begin{figure}[h!tb]
\begin{center}
\includegraphics[height=11.cm]{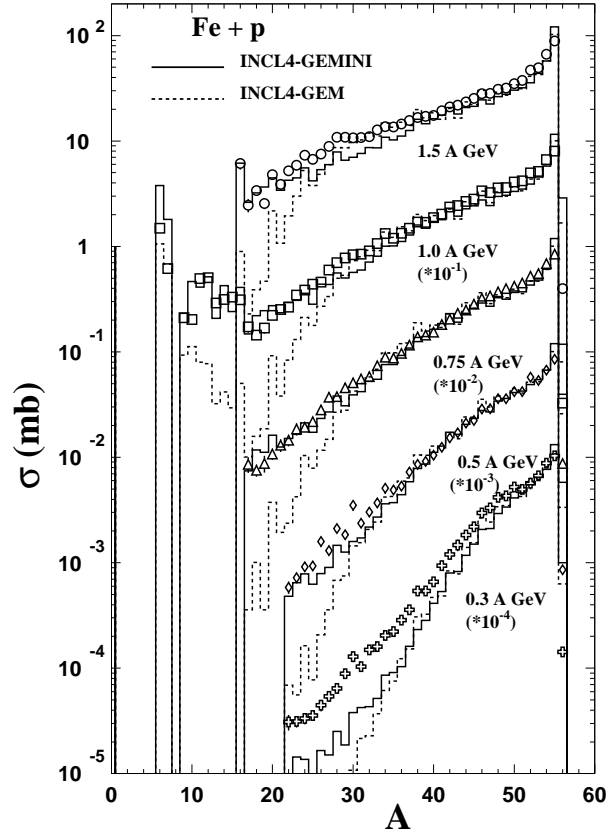}
\end{center}
\caption{{Spallation residue cross-sections of iron as a function of there mass number compared with a calculation with INCL4 coupled with GEM (dashed lines) or GEMINI (continuous lines). Points are data of the present paper complemented for low masses at 1 GeV by the ones of ~\cite{Paolo} obtained during the same experiment.}}
\label{fig:gem-gemini}
\end{figure}

{Figure}~\ref{fig:gem-gemini} shows the INCL4 intranuclear cascade coupled with the GEM model, which takes into account also evaporation of intermediate-mass fragments. The calculated cross-sections for the intermediate mass residues are improved comparatively to ABLA. However one observes a slight underestimation of the residues close to iron and the underprediction of the very light fragments still persists for masses slightly smaller than with ABLA. 

From the comparison with the three evaporation models (Dresner, ABLA and GEM) and the remark concerning excitation energy made above, it can be presumed that standard evaporation models, even including the emission of IMF (GEM), cannot reproduce the bulk of our data. This is why we tried other models which include other de-excitation modes.

On {Fig.}~\ref{fig:gem-gemini}, are also shown the predictions of GEMINI. If on the heavy fragments the results are slightly less
satisfactory than with GEM, the behaviour for A lower than 30 is significantly improved. Probably due to its capability of predicting asymmetric fission in the Transition State Model prescription, GEMINI appears as the best suited code to reproduce the bulk of the data except at the lower energy (300 MeV per nucleon).  Actually at 300 MeV per nucleon, all the calculations, whatever the choice of INC or de-excitation models, start to deviate from experiment around A equal 48.

\begin{figure}[h!tb]
\begin{minipage}[c]{17.4cm}
\begin{center}
\includegraphics[width=10cm]{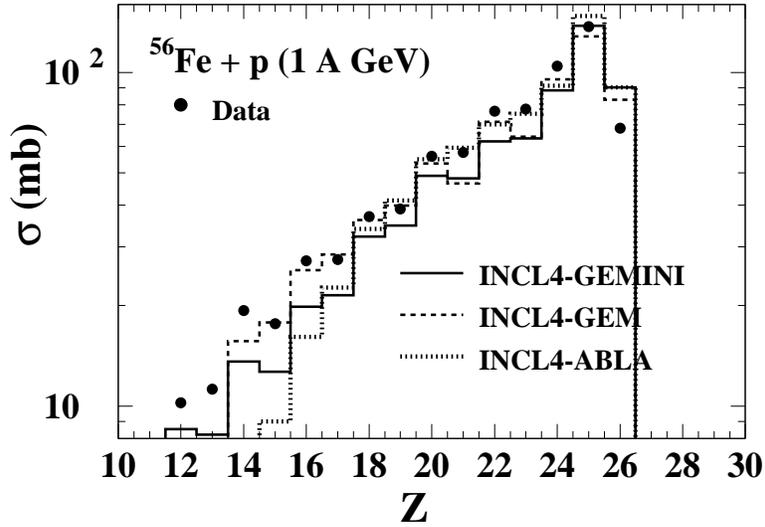}
\end{center}
\caption{{Charge distribution of the spallation residues of iron at 1 A~GeV compared to the INCL4 cascade coupled to the deexcitation models GEM, GEMINI and ABLA.}}
\label{fig:zoddeven}
\end{minipage}
\end{figure}

Even-odd disymmetry of the cross sections are clearly visible on an unlarged picture of the Z distribution at 1 GeV per nucleon in 
{Fig.}~\ref{fig:zoddeven} representative also of other energies. In spite of a small underprediction of the absolute cross-sections with GEMINI, the ratios between odd and even Z cross-sections are very close to the experimental ones. Whereas GEM gives a too strong effect, ABLA predicts (with the present version) a slightly too small even-odd effect. But for the largest cross-sections above 18 the INCL4-ABLA remains the more precise prediction of the experimental values.  

\begin{figure}[h!tb]
\begin{center}
\includegraphics[width=8.6cm]{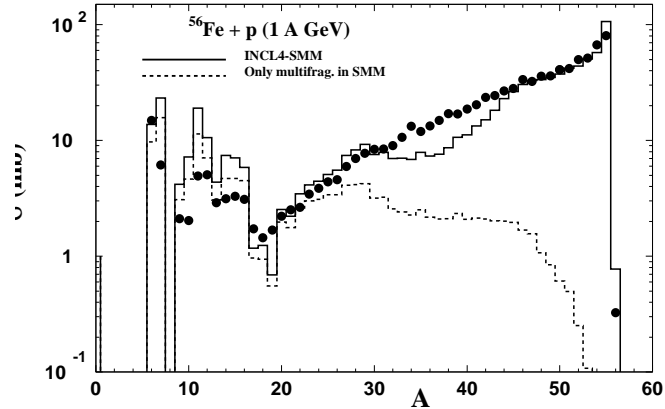}
\end{center}
\caption{{Calculation of the cross sections with INCL4 and SMM (solid line) compared to data points at 1 A~GeV. The multifragmentation
contribution to the calculation is the dashed line.}}
\label{fig:smm_multifrag}
\end{figure}

Another mechanism that could be invoked to explain our large yields of light fragments could be the onset of multifragmentation at the highest excitation energies~\cite{Paolo}. The coupling of INCL4 with the multifragmentation model, SMM, is shown in {Fig.}~\ref{fig:smm_multifrag}. The model well describes the heavy residues and the ones with mass between 20 and 30. However, it overpredicts the lighest fragments and disagrees strongly with the data in the A region 30-45. The contribution of fragments produced by multifragmentation  is shown as the dashed curve in the figure (multifragmentation events being identified by the entry into the multifragmentation routine in the code~\cite{Bot}). The major part of the light fragment cross-section is produced by multifragmentation while masses above 25 are mostly originating from evaporation. However, it is likely the opening of multifragmentation that causes the hole in the region $A=30-45$, not observed experimentally. Our results are at variance with what was found in~\cite{Paolo}, where SMM coupled to another INC model (from~\cite{cem}) was giving a good agreement with the data, provided that a pre-equilibrium stage was added. With INCL4, which as explained in~\cite{INCL4}, handles what is often called the pre-equilibrium stage, the best agreement with the whole set of data is obtained with GEMINI.

\bigskip

However, a clear conclusion on the mechanism responsible for the light and intermediate fragment production is difficult and would need more constraining information. It seems rather clear that the traditional Weisskopf-Ewing evaporation as used in ABLA or even in GEM, which evaporates IMFs, miss the production of the lighest nuclei. However, the reason for the success of GEMINI, Hauser-Fesbach treatment or asymmetric fission from the Transition-state-model, is not fully understood and a possible contribution of multifragmentation is not ruled out. Forthcoming exclusive experiments will probably help to clarify the situation by an identification of the various fragments emitted in coincidence during the de-excitation stage of the reaction.

 \subsubsection{Isotopic distributions}

\begin{figure}[h!tb]
\begin{center}
\includegraphics[width=10.0cm,height=8.0cm]{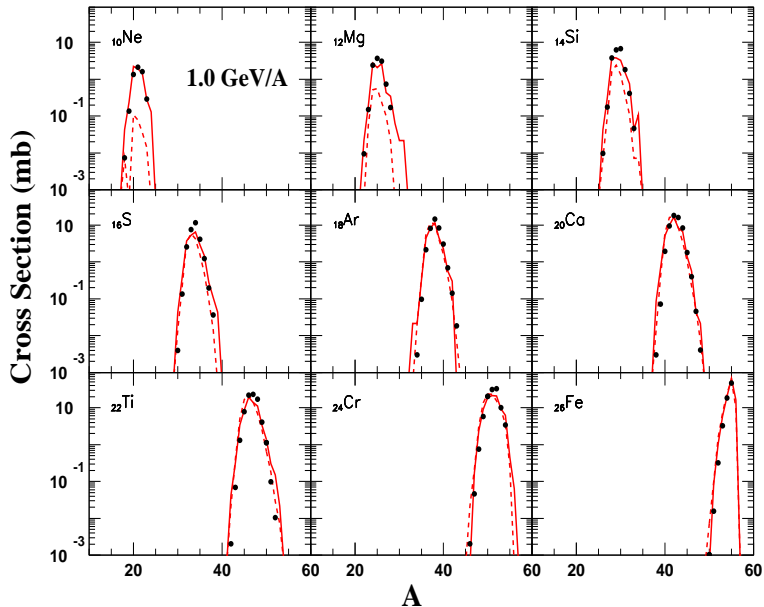}
\end{center}
\caption{{A selection of isotopic distributions of cross-sections measured at 1 A~GeV compared with INCL4-GEMINI calculations (continuous lines) and INCL4-ABLA (dashed lines).}}
\label{fig:bof}
\end{figure}

In this experiment, more than 500 individual isotopic cross sections have been measured which have been compared systematically to calculations done with the four different de-excitation codes (ABLA, GEM, GEMINI and SMM) coupled with INCL4. As an example, the comparison of GEMINI (full line) and ABLA (dashed line) with a selection of measured isotopic cross sections at 1 A~GeV is shown on {Fig.}~\ref{fig:bof}. Except for the better level of cross sections for light residues from GEMINI, already seen when looking at the mass distributions, it is difficult to conclude about the detailed quality of each model.

\begin{figure}[h!tb]
\begin{center}
\includegraphics[height=15.0cm]{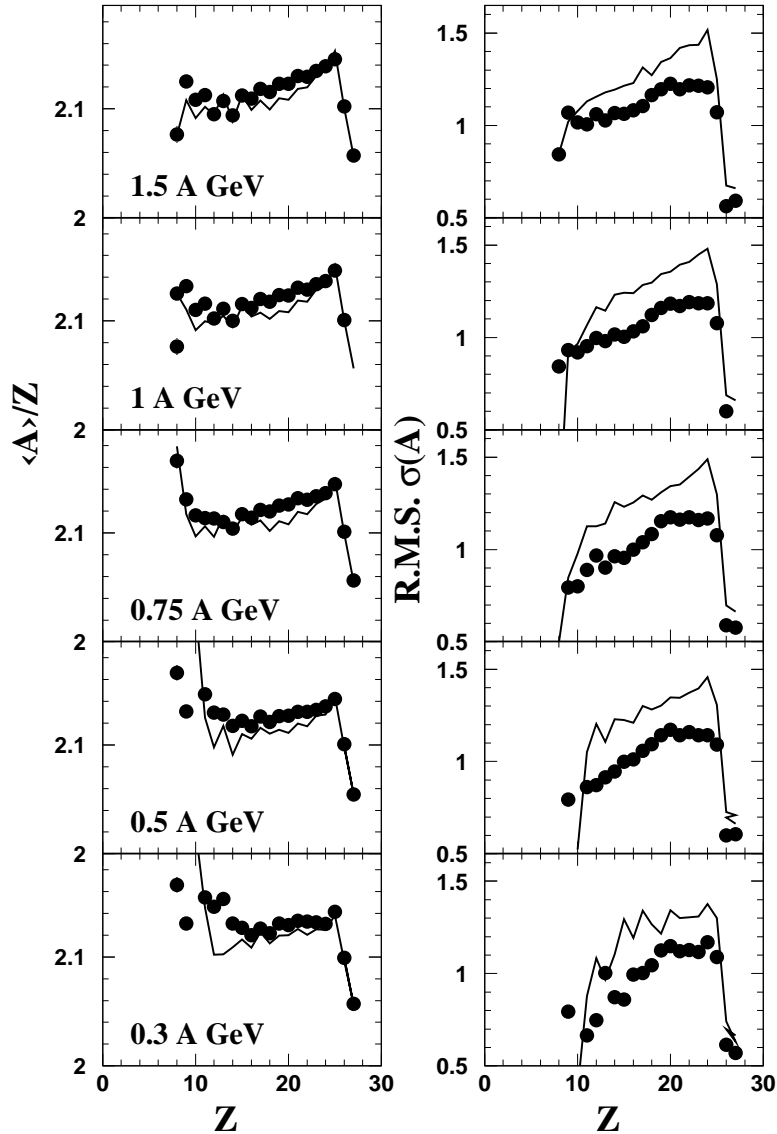}
\end{center}
\caption{{Mean atomic mass A over Z ($\langle A\rangle/Z$) and rms ($\sigma(A)$) of the isotopic cross section distributions as a function of the atomic number Z, at the five bombarding energies, compared to calculations done with the intranuclear cascade INCL4 model coupled with GEMINI. The calculated values have been averaged over the actually measured isotopes.}}
\label{fig:amean_energy}
\end{figure}

\begin{figure}[h!tb]
\begin{center}
\includegraphics[height=15.0cm]{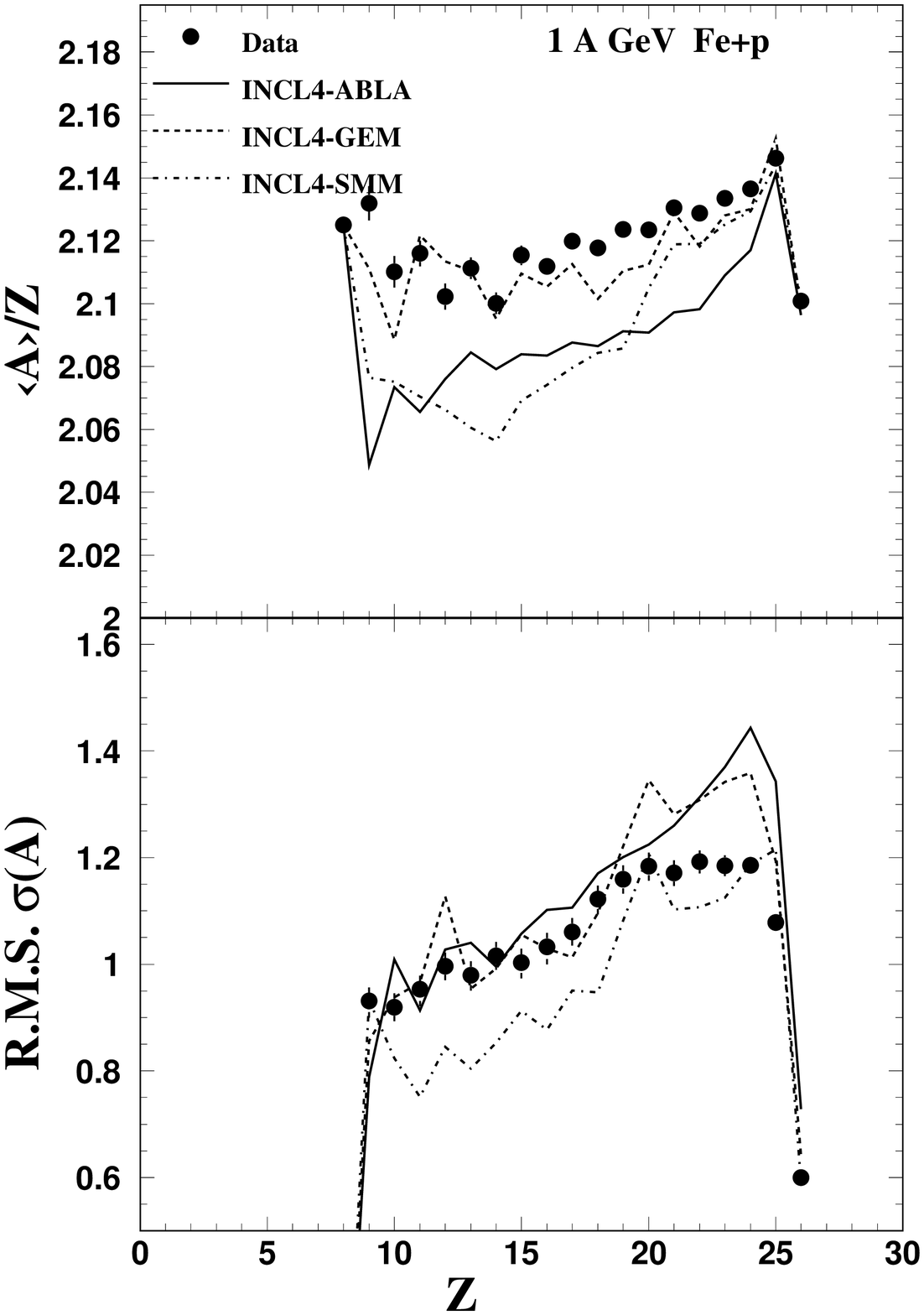}
\end{center}
\caption{{Same as fig~\ref{fig:amean_energy} but only at 1 A~GeV for comparison with calculations done with the intranuclear-cascade INCL4 model coupled with ABLA (full line), GEM (dashed line), and SMM (dashed-dotted line).}}
\label{fig:amean_model}
\end{figure}

A more powerfull way to make the comparison is to look at the shapes of the isotopic distributions for each element through the mean atomic mass ($\langle A\rangle$) divided by the charge Z of the element and the width (root mean square) of the measured (or computed) distributions. {Figure}~\ref{fig:amean_energy} presents the comparison of these quantities with GEMINI at the five energies while {Fig.}~\ref{fig:amean_model} shows the results INCL4 coupled to either ABLA, GEM or SMM at 1 A~GeV. Actually, it is remarkable that the deviations between models and experiment are qualitatively independent of the beam energy. This can be checked on {Fig.}~\ref{fig:amean_energy} for GEMINI but holds also for the comparison with the other models. For this reason, the comparison with the other three models is shown only at 1 A~GeV in {Fig.}~\ref{fig:amean_model}. But again, the following conclusions are the same at all the energies.

For Z equal 25, 26 and 27 (not measured at 1 A~GeV), cross sections are dominated by the cascade, leaving the remnant nucleus with very little excitation energy. Therefore, the choice of the evaporation model play practically no role and basically the $\langle A\rangle/Z$ is perfectly reproduced.
The average value of the isotopic distribution ($\langle A\rangle/Z$) is actually very well predicted by GEM and GEMINI on all the range (down to Z equal 8 to 9), with the correct odd-even effects, whereas ABLA gives a value systematically too small. The SMM model gives a correct centroid down to Z equal 20 but is the worst below this value with a distribution centered one mass below the data at lower Z. As regards the width of the distributions, none of the models is good on all the Z range. The widths computed from GEMINI are systematically a little too wide.  With GEM and ABLA, they are too wide only in the range 20-25, otherwise they are very close to the data. For SMM, it is the contrary, rather good at high Z but too narrow for lower charges. This fact was already noticed in~\cite{Souza}.

All this shows that none of the de-excitation models is perfect. However, taking into account the information from both the cross-sections and the isotopic distribution shapes, it can be concluded that the GEMINI gives the best agreement with our data.

 \subsubsection{Recoil velocities.}

\begin{figure}[h!tb]
\begin{center}
\includegraphics[width=5.0cm,angle=-90]{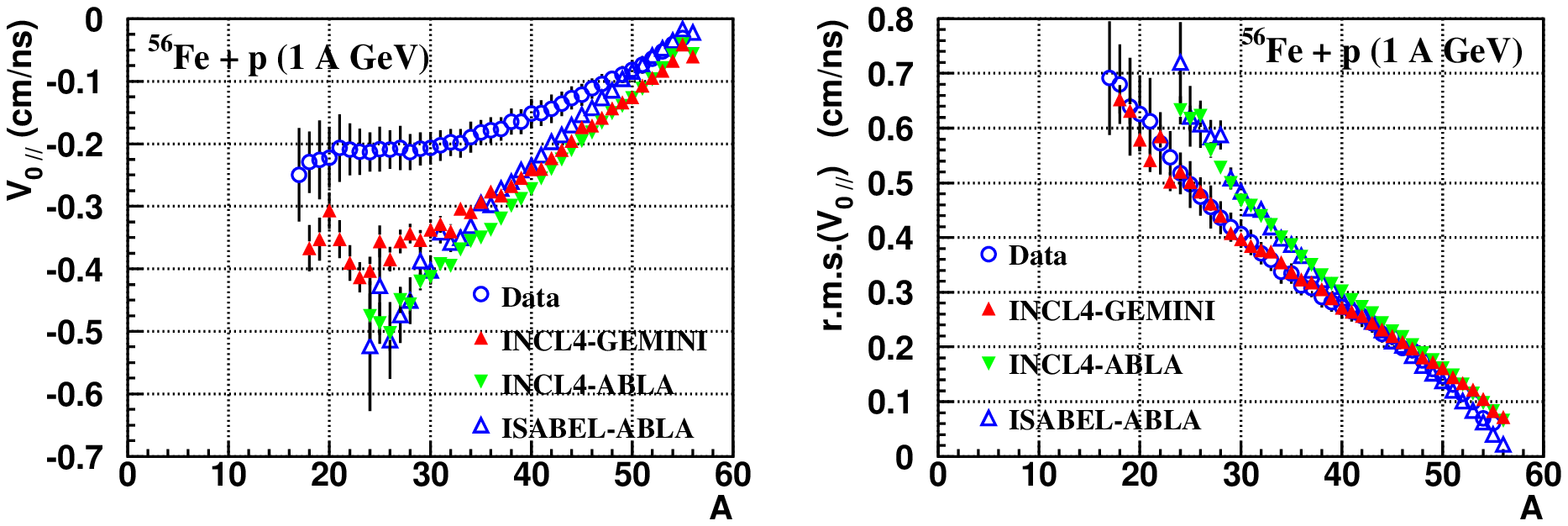}
\end{center}
\caption{{Mean (left) and Root Mean Square (right) values of the longitudinal recoil velocity distribution for spallation residues at 1 A~GeV versus their atomic mass. Open circles are the experimental values. Down triangles are predictions from the INCL4-ABLA calculation, up empty triangles from the ISABEL-ABLA and up full triangles from the INCL4-GEMINI calculation. Velocities are expressed in the beam ($^{56}Fe$) rest frame and with a minus sign as being opposite to the iron beam direction.}}
\label{fig:v}
\end{figure}

Concerning the kinetic characteristics of the fragments, we show in {Fig.}~\ref{fig:v} a comparison between the experimental mean longitudinal recoil velocities for each mass and the ones calculated with the INCL4 model combined with ABLA or GEMINI at 1 A~GeV. The same comparison is also done for the width (root mean square) of the longitudinal distribution (right part of the figure).

\begin{figure}[h!tb]
\begin{center}
\includegraphics[height=13.0cm,angle=-90]{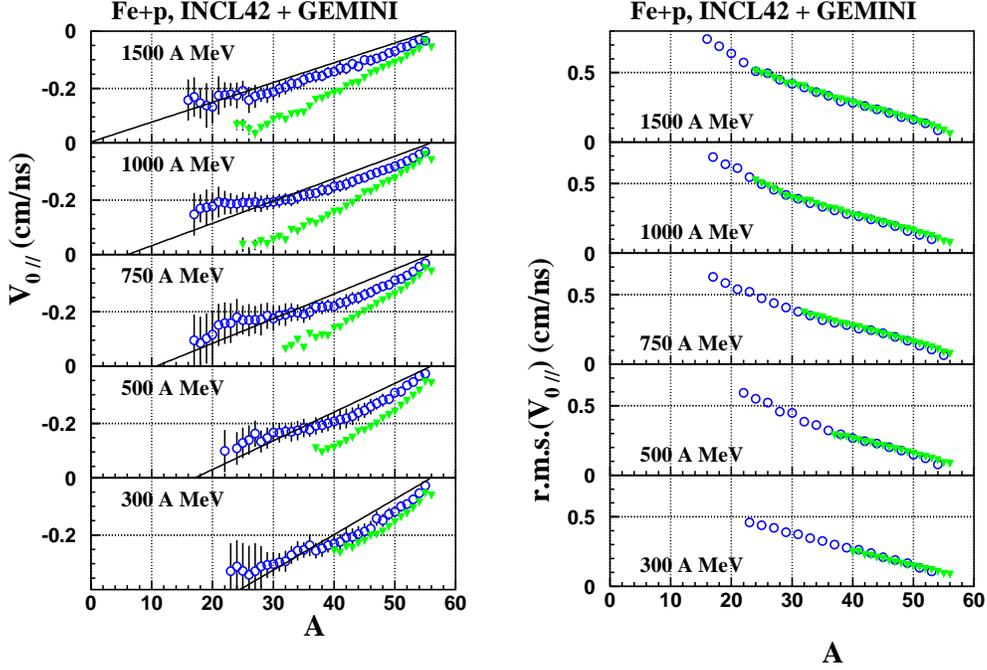}
\end{center}
\caption{{Mean values and r.m.s. of the longitudinal recoil velocity distribution for spallation residues versus their atomic mass at all beam
energies. Open circles are experimental values. Down triangles are predictions from the INCL4 + GEMINI calculation. The lines are the Morrissey systematics~\cite{{Morri}}.}}
\label{fig:all_speed}
\end{figure}

One can observe an important discrepancy between the experimental mean recoil velocities and values predicted by the models. It is worthwhile to note that the experimental data decrease much more slowly with decreasing mass than the values predicted by the codes. Furthermore, they seem to saturate at a mass value of 35. This saturation is not seen with ABLA. Only GEMINI shows a clear tendency towards saturation below A=30. For the widths, on the contrary, the agreement with the experimental data is better, especially when using GEMINI. This behavior, presented here at 1 GeV, is very similar at the others energies analyzed in this experiment ({Fig.}~\ref{fig:all_speed}). The better agreement with GEMINI could be due to the existence of binary decays in de-excitation phase that reduces the mean longitudinal velocity of the final residual nuclei since the recoil momentum will originate from a heavier nucleus and will be split between two partners emitted in an arbitrary direction with respect to the beam one. In the same figure, are also shown the predictions from systematics of Morrisey~\cite{Morri}, which more or less give the correct slope for large mass but miss the saturation observed in the data. Actually, the two other de-excitation models GEM and SMM, not shown here, give results rather similar to ABLA: rather good for the widths but a slope too steep and an inability to describe the saturation of the mean values.

The fact that the mean recoil velocities for the heaviest masses is not well predicted cannot be ascribed to the de-excitation models but should rather raise questions on the intranuclear cascade. This is why we have also performed a calculation using ISABEL coupled to ABLA, which is presented in {Fig.}~\ref{fig:v}. Obviously, ISABEL better reproduces both the mean values and the widths for masses larger than 50, indicating a possible deficiency of INCL4 in the recoil velocity determination. Actually, a similar systematic deviation of INCL4 concerning the mean velocities has already been noticed for Pb+p at 1 GeV/A~\cite{Wla}. However, the general trend of the ISABEL calculation on the whole mass range leads to the same conclusion that it is unable to give the correct slope and saturation effect of the experimental data.

\bigskip

\section{Conclusion}

The spallation residues produced in the bombardment of $^{56}Fe$ at 1.5 , 1.0, 0.75, 0.5 and 0.3 A~GeV on a liquid-hydrogen target have been studied using the reverse kinematics technique and the Fragment Separator at GSI (Darmstadt). This technique has permitted the full identification in charge and mass of all isotopes produced with cross-sections larger than $10^{-2}$ mb down to $Z=8$. Their individual production cross-sections and recoil velocities at the five energies have been obtained. 

The production cross-sections have been compared with the previously existing data, either charge-changing cross-sections with a few isotopic cross-sections at one energy measured in reverse kinematics or excitation functions for a limited number of isotopes obtained by $\gamma$-spectrometry in direct kinematics. Globally, our results were found in good agreement with former data. This comparison also showed that our experimental method leads to a much more complete picture of the residue production than what was possible before with the few scattered results, allowing sometimes to detect possible inconsistencies in other sets of data. 

Comparisons with parametric formulas, often used in astrophysics, have been performed: the Webber formula gives rather good predictions of the charge distributions but produces too narrow isotopic distributions. It also totally fails for the lightest nuclei (below A equal 30-35) in the region not measured at the time when this formula was derived. The EPAX formula (once renormalized to give the correct total reaction cross-section) is usable only in the limiting fragmentation regime, apparently not yet fully reached at 1.5 A~GeV. However, it nearly gives the right A dependence of the cross sections at our highest energy. The best formula seems to be the Silberberg and Tsao one, which is in very good agreement with the experimental mass distributions and mean value of the isotopic distributions at all the energies except at 300 MeV (as all the models). The use of parametric formulas can be of great help for a fast estimation for certain applications, but the example of Webber's illustrates the possible danger of using parametric formulas outside the range on which they have been adjusted. Our data could certainly be used to derive new, more relaible parametric formulas for use in cosmic-ray propagation codes. 

Predictions of different intranuclear-cascade models (Bertini, ISABEL and INCL4) combined with different de-excitation models (Dresner, ABLA, GEM, SMM and GEMINI) have been confronted to the new experimental data. INCL4 or ISABEL combined with standard Weisskopf-Ewing evaporation models as ABLA or GEM give a good description of the residual production close in charge to iron but they underpredict systematically the light evaporation residues in the mass region 20-30. This fact, together with the saturation observed in the experimental longitudinal velocity at low masses, could be an indication that another de-excitation mechanism has to be considered. A de-excitation including a possible contribution from multifragmentation, as treated by SMM, improves significantly the predictions of light and intermediate mass fragments but at the detriment of residues in the region $A=30-45$. SMM also misses the saturation of the recoil velocity and do not properly predict the isotopic distribution mean values and widths. The best overall agreement with the data is obtained with GEMINI. This GEMINI model gives a rather precise account of all cross sections measured here as a function of the beam energy. The recoil velocities, although not perfect, are the closest to the experimental values and the mean values and widths of the isotopic distributions are rather well reproduced. Other authors~\cite{NESSI, NESSI2} have found that generally GEMINI reproduces very well the energy spectra of both light charged particles and intermediate mass fragments in a wide range of incident energies and target masses. Similar conclusions (best agreement with GEMINI) have been reached by~\cite{maschnik_santafe} using as intra-nuclear model the Cascade-Exciton Model, coupled with GEM, GEMINI and SMM and compared with these data taken from the C. Villagrasa-Canton PhD ~\cite{cv_these}. In~\cite{Paolo}, with another INC coupled with a preequilibrium stage, the deexcitation code SMM was found to give the best agreement with the 1 GeV data. It is obviously difficult to definitively conclude on the production mechanism of the intermediate and light mass fragments and probably only additional experimental information on correlations between residual nuclei and light particles could bring aswers to the questions addressed here.

As regards to the potential interest of the present data for applications, we supply isotopic cross-sections that can be used to directly estimate the change in chemical composition that could occur in an ADS window made predominantly of iron and recoil velocities to calculate damages due to atomic displacements (DPA)~\cite{cv_these} . 

\section{Acknowledgements}
This experiment has benefited from a flexible and efficient driving of the GSI accelerators. Many thanks to all the driving team. The technical support of K.H. Behr, A. Br\"unle and K. Burkard was crucial for the preparation of the experimental setup, and the liquid target was smoothly managed by P. Chesny, J.M. Gheller and G. Guiller.  We thank them all for their contribution.
Comparison with irradiation experiments (fig 14 and 15) was possible due to the help of J.C. David and due to the kindness of Pr. R. Michel providing us the experimental file.
We thank also A. Botvina for helping us to identify the multifragmentation contribution in the SMM code. 

During the interpretation of these results, we have benefited from fruitful discussions with J. Cugnon and Y. Yariv. 

\newpage
\include{anexe_sec}

\newpage
\include{anexe_v}

\end{document}

%% file: anexe_sec.tex

\section{Appendix A: Cross sections.\label{annexe_xs}}

    \begin{table}[h!]
    \centering
\begin{tabular}{ccccccccc}
    & & 1500 A MeV & & & $\sigma_R(mb) = 822\pm73 $ & \\
\hline 
\hline   
 Z &   A &     $\sigma(mb)$      &     Z &   A &     $\sigma(mb)$    &    Z &   A &     $\sigma(mb)$   \\\hline\hline
 27&   54&   0.035$\pm$0.0021& 23&  50&   16.272$\pm$2.383& 20&  44& 8.0592$\pm$0.798  \\\hline
 27&   55&   0.245$\pm$0.0104& 23&  51&    7.179$\pm$0.446& 20&  45& 1.8149$\pm$0.174  \\\hline
 27&   56&   0.413$\pm$0.0187& 23&  52&    1.274$\pm$0.073& 20&  46& 0.4263$\pm$0.039  \\\hline
 26&   51&   0.016$\pm$0.0021& 23&  53&    0.251$\pm$0.012& 20&  47& 0.0497$\pm$0.004  \\\hline
 26&   52&   0.309$\pm$0.0187& 23&  54&    0.004$\pm$0.001& 20&  48& 0.0052$\pm$0.001  \\\hline
 26&   53&   2.865$\pm$0.1338& 22&  43&    0.093$\pm$0.009& 19&  37& 0.0683$\pm$0.008  \\
\hline
\hline   
							 
 \end{tabular}
  \end{table}

\clearpage
    
    \begin{table}
    \centering 
    \begin{tabular}{ccccccccc}
    & & 1500 A MeV & & & & & \\ 
\hline 
\hline
     	  Z &	A &	$\sigma(mb)$  & Z &   A &     $\sigma(mb)$      &      Z &   A &	    $\sigma(mb)$  \\\hline\hline	       
 17&   38&   1.872$\pm$0.2066& 14&  29&    8.920$\pm$1.055& 11&  25& 1.0696$\pm$0.130  \\\hline
 17&   39&   0.536$\pm$0.0589& 14&  30&    8.572$\pm$1.042& 11&  26& 0.1719$\pm$0.021  \\\hline
 17&   40&   0.098$\pm$0.0103& 14&  31&    2.549$\pm$0.314& 11&  27& 0.0309$\pm$0.003  \\\hline
 17&   41&   0.018$\pm$0.0021& 14&  32&    0.626$\pm$0.073& 10&  19& 0.2531$\pm$0.032  \\\hline
 16&   31&   0.027$\pm$0.0031& 14&  33&    0.072$\pm$0.008& 10&  20& 3.2281$\pm$0.402  \\\hline
 16&   32&   3.679$\pm$0.4306& 14&  34&    0.015$\pm$0.002& 10&  21& 3.1828$\pm$0.446  \\
\hline          	 
  \hline
    							   
 \end{tabular}
  \end{table}

\clearpage
    
    \begin{table}
    \centering 
    \begin{tabular}{ccccccccc}
    & & 1000 A MeV & & & $\sigma_R(mb) = 811\pm76 $ & & \\
\hline 
\hline   
    Z &   A &     $\sigma(mb)$      &      Z &   A &	   $\sigma(mb)$   &  Z &   A &     $\sigma(mb)$   \\\hline\hline
 26&   51&   0.016$\pm$0.0010& 23&  52&    1.259$\pm$0.071& 20&  44& 8.6501$\pm$0.896  \\\hline
 26&   52&   0.333$\pm$0.0187& 23&  53&    0.246$\pm$0.011& 20&  45& 1.8676$\pm$0.178  \\\hline
 26&   53&   3.319$\pm$0.1556& 23&  54&    0.004$\pm$0.001& 20&  46& 0.4159$\pm$0.038  \\\hline
 26&   54&  18.977$\pm$0.7686& 22&  42&    0.002$\pm$0.001& 20&  47& 0.0476$\pm$0.004  \\\hline
 26&   55&  47.982$\pm$1.7913& 22&  43&    0.071$\pm$0.007& 20&  48& 0.0041$\pm$0.001  \\\hline
 25&   49&   0.017$\pm$0.0010& 22&  44&    1.347$\pm$0.134& 19&  37& 0.0455$\pm$0.005  \\
\hline
\hline							 
 \end{tabular}
  \end{table}

\clearpage

    \begin{table}
    \centering 
    \begin{tabular}{cccccc}
    & &  1000 A MeV & & \\
\hline 
\hline   
    Z &   A &     $\sigma(mb)$      &      Z &   A &	   $\sigma(mb)$     \\\hline\hline
 17&   38&   1.608$\pm$0.1767& 13&  27&    5.202$\pm$0.625  \\\hline
 17&   39&   0.464$\pm$0.0506& 13&  28&    3.173$\pm$0.384  \\\hline
 17&   40&   0.080$\pm$0.0083& 13&  29&    1.393$\pm$0.171  \\\hline
 16&   30&   0.004$\pm$0.0010& 13&  30&    0.229$\pm$0.027  \\\hline
 16&   31&   0.138$\pm$0.0165& 13&  31&    0.047$\pm$0.005  \\\hline
 16&   32&   2.611$\pm$0.3056& 12&  22&    0.010$\pm$0.001  \\
\hline
    
 \end{tabular}
  \end{table}

\clearpage

    \begin{table}
    \centering 
    \begin{tabular}{ccccccccc}
    & & 750 A MeV & & & $\sigma_R(mb) = 767\pm66 $ & & \\
\hline 
\hline   
    Z &   A &     $\sigma(mb)$      &      Z &   A &	   $\sigma(mb)$   &  Z &	A &	$\sigma(mb)$   \\\hline\hline
 27&   54&   0.035$\pm$0.0021& 23&  50&   19.689$\pm$2.487& 20&  44& 8.3324$\pm$0.831  \\\hline
 27&   55&   0.333$\pm$0.0145& 23&  51&    7.647$\pm$0.473& 20&  45& 1.8625$\pm$0.178  \\\hline
 27&   56&   0.474$\pm$0.0228& 23&  52&    1.270$\pm$0.073& 20&  46& 0.3901$\pm$0.036  \\\hline
 26&   51&   0.011$\pm$0.0010& 23&  53&    0.225$\pm$0.010& 20&  47& 0.0445$\pm$0.004  \\\hline
 26&   52&   0.300$\pm$0.0176& 23&  54&    0.003$\pm$0.001& 20&  48& 0.0041$\pm$0.001  \\\hline
 26&   53&   3.437$\pm$0.1628& 22&  43&    0.045$\pm$0.004& 19&  37& 0.0207$\pm$0.002  \\
\hline
\hline							 		     
 \end{tabular}
  \end{table}

\clearpage

    \begin{table}
    \centering
 
    \begin{tabular}{cccccc}
    & &  750 A MeV & &  \\
\hline 
\hline   
    Z &   A &     $\sigma(mb)$      &      Z &   A &	   $\sigma(mb)$  \\\hline\hline
 17&   39&   0.322$\pm$0.0351& 13&  28&    1.841$\pm$0.225  \\\hline
 17&   40&   0.052$\pm$0.0052& 13&  29&    0.714$\pm$0.087  \\\hline
 16&   31&   0.046$\pm$0.0062& 13&  30&    0.088$\pm$0.010  \\\hline
 16&   32&   1.437$\pm$0.1683& 13&  31&    0.020$\pm$0.002  \\\hline
 16&   33&   5.539$\pm$0.6412& 12&  23&    0.052$\pm$0.008  \\\hline
 16&   34&   8.110$\pm$0.9386& 12&  24&    1.193$\pm$0.147  \\
\hline
\hline							 		      
 \end{tabular}
  \end{table}

\clearpage

    \begin{table}
    \centering
 
    \begin{tabular}{ccccccccc} 
    & & 500 A MeV & & & $\sigma_R(mb) = 660\pm53 $ & & \\
\hline 
\hline   
    Z &   A &     $\sigma(mb)$      &      Z &   A &       $\sigma(mb)$   &  Z &   A &     $\sigma(mb)$   \\\hline\hline
 27&   53&   0.003$\pm$0.0010& 23&  49&   27.507$\pm$2.593& 20&  43&11.4146$\pm$1.372  \\\hline
 27&   54&   0.059$\pm$0.0021& 23&  50&   19.027$\pm$2.696& 20&  44& 6.7186$\pm$0.745  \\\hline
 27&   55&   0.485$\pm$0.0197& 23&  51&    6.266$\pm$0.387& 20&  45& 1.2635$\pm$0.121  \\\hline
 27&   56&   0.611$\pm$0.0248& 23&  52&    0.949$\pm$0.054& 20&  46& 0.2558$\pm$0.024  \\\hline
 26&   51&   0.020$\pm$0.0010& 23&  53&    0.152$\pm$0.007& 20&  47& 0.0268$\pm$0.002  \\\hline
 26&   52&   0.416$\pm$0.0238& 23&  54&    0.002$\pm$0.001& 20&  48& 0.0021$\pm$0.001  \\
\hline
\hline							 		      
 \end{tabular}
  \end{table}

\clearpage

    \begin{table}
    \centering
 
    \begin{tabular}{cccccc}
    & &  500 A MeV & & \\
\hline 
\hline   
    Z &   A &     $\sigma(mb)$      &      Z &   A &       $\sigma(mb)$    \\\hline\hline
 17&   39&   0.188$\pm$0.0206& 13&  30&    0.083$\pm$0.010  \\\hline
 17&   40&   0.028$\pm$0.0031& 13&  31&    0.009$\pm$0.001  \\\hline
 16&   31&   0.031$\pm$0.0041& 12&  23&    0.009$\pm$0.002  \\\hline
 16&   32&   0.822$\pm$0.0968& 12&  24&    0.354$\pm$0.044  \\\hline
 16&   33&   2.686$\pm$0.3109& 12&  25&    0.748$\pm$0.093  \\\hline
 16&   34&   4.570$\pm$0.5764& 12&  26&    1.359$\pm$0.188  \\
\hline
\hline							 		      
 \end{tabular}
  \end{table}

\clearpage

    \begin{table}
    \centering
 
    \begin{tabular}{ccccccccc}
    & & 300 A MeV & & & $\sigma_R(mb) = 701\pm56 $ & & \\
\hline 
\hline   
    Z &   A &     $\sigma(mb)$      &      Z &   A &       $\sigma(mb)$   &  Z &   A &     $\sigma(mb)$   \\\hline\hline
 27&   54&   0.092$\pm$0.0041& 23&  50&   22.319$\pm$2.417& 20&  44& 5.7367$\pm$0.636  \\\hline
 27&   55&   0.913$\pm$0.0341& 23&  51&    6.935$\pm$0.429& 20&  45& 0.9860$\pm$0.094  \\\hline
 27&   56&   1.390$\pm$0.0538& 23&  52&    1.014$\pm$0.057& 20&  46& 0.2063$\pm$0.020  \\\hline
 26&   51&   0.021$\pm$0.0010& 23&  53&    0.133$\pm$0.006& 20&  47& 0.0196$\pm$0.002  \\\hline
 26&   52&   0.554$\pm$0.0310& 23&  54&    0.001$\pm$0.001& 19&  37& 0.0113$\pm$0.001  \\\hline
 26&   53&   5.305$\pm$0.2462& 22&  43&    0.039$\pm$0.004& 19&  38& 0.2608$\pm$0.029  \\
\hline
\hline							 		      
 \end{tabular}
  \end{table}

\clearpage

    \begin{table}
    \centering
 
    \begin{tabular}{cccccc}
    & &  300 A MeV & &  \\
\hline 
\hline   
    Z &   A &     $\sigma(mb)$      &      Z &   A &       $\sigma(mb)$     \\\hline\hline
 17&   41&   0.002$\pm$0.0010& 14&  31&    0.386$\pm$0.051  \\\hline
 16&   31&   0.006$\pm$0.0010& 14&  32&    0.047$\pm$0.006  \\\hline
 16&   32&   0.271$\pm$0.0319& 14&  33&    0.005$\pm$0.001  \\\hline
 16&   33&   1.199$\pm$0.1390& 13&  26&    0.059$\pm$0.007  \\\hline
 16&   34&   1.888$\pm$0.2213& 13&  27&    0.438$\pm$0.059  \\\hline
 16&   35&   0.827$\pm$0.1050& 13&  28&    0.491$\pm$0.067  \\\hline
 16&   36&   0.239$\pm$0.0268& 13&  29&    0.334$\pm$0.044  \\\hline
 16&   37&   0.027$\pm$0.0031& 13&  30&    0.098$\pm$0.013  \\\hline
 16&   38&   0.005$\pm$0.0010& 13&  31&    0.004$\pm$0.001  \\\hline
 15&   29&   0.007$\pm$0.0010& 12&  24&    0.023$\pm$0.004  \\\hline
 15&   30&   0.081$\pm$0.0103& 12&  25&    0.288$\pm$0.037  \\\hline
 15&   31&   0.670$\pm$0.0802& 12&  26&    0.399$\pm$0.056  \\\hline
 15&   32&   1.216$\pm$0.1574& 12&  27&    0.120$\pm$0.016  \\\hline
 15&   33&   0.443$\pm$0.0576& 12&  28&    0.004$\pm$0.001  \\\hline
 15&   34&   0.062$\pm$0.0072& 11&  23&    0.280$\pm$0.040  \\\hline
 15&   35&   0.010$\pm$0.0010& 11&  24&    0.319$\pm$0.041  \\\hline
 14&   28&   0.165$\pm$0.0195& 11&  25&    0.078$\pm$0.010  \\\hline
 14&   29&   0.571$\pm$0.0689& 10&  22&    0.316$\pm$0.053  \\\hline
 14&   30&   1.146$\pm$0.1522& 10&  23&    0.041$\pm$0.006  \\\hline
    \hline
							 		      
 \end{tabular}
  \end{table}

\clearpage

%% file: anexe_v.tex
    \section{Appendix B: Longitudinal velocity of residual nuclei (Mean and R.M.S. value) in the iron beam system at rest.\label{annexe_v}}

    \begin{table}[h!]
    \centering
    \begin{tabular}{cccccccc} 
    & & & 1500 A MeV & & & & I= \mbox{Interpolated}\\  
\hline 
\hline   
    Z &   A &     Mean velocity(cm/ns)  & R.M.S. (cm/ns)    &      Z &   A &	  Mean velocity(cm/ns)    &   R.M.S. (cm/ns) \\\hline\hline
    27 &  54   & -0.079$\pm$0.020        & 0.127 & 22      & 47   &-0.098(I)      &0.198		   \\\hline
    27 &  55   & -0.058$\pm$0.015 	     & 0.110 & 22      & 48   &-0.089(I)      &0.176	       \\\hline
    27 &  56   & -0.033$\pm$0.008 	     & 0.104 & 22      & 49   &-0.086$\pm$0.017   &0.172	       \\\hline
    26 &  51   & -0.093$\pm$0.023 	     & 0.141 & 22      & 50   &-0.074$\pm$0.015   &0.161		\\\hline
    26 &  52   & -0.056$\pm$0.014 	     & 0.132 & 22      & 51   &-0.075$\pm$0.015   &0.165	       \\\hline
    26 &  53   & -0.035$\pm$0.009 	     & 0.114 & 22      & 52   &-0.056$\pm$0.017   &0.138	       \\
\hline
    \hline

 \end{tabular}
  \end{table}

\clearpage

    \begin{table}
    \centering
    \begin{tabular}{cccccccc}
    & & & 1500 A MeV & & & &I= \mbox{Interpolated} \\ 
\hline 
\hline   
    Z &   A &     Mean velocity(cm/ns)  & R.M.S. (cm/ns)    &      Z &   A &	  Mean velocity(cm/ns)    &   R.M.S. (cm/ns) \\\hline\hline
     18 &   36 &-0.164$\pm$0.025  & 0.342 &  13 &26 &-0.242$\pm$0.048  & 0.510        \\\hline
     18 &   37 &-0.156$\pm$0.023  & 0.322 &  13 &27 &-0.232$\pm$0.046  & 0.467    \\\hline
     18 &   38 &-0.158$\pm$0.024  & 0.291 &  13 &28 &-0.221$\pm$0.044  & 0.449    \\\hline
     18 &   39 &-0.148(I)     & 0.285 &  13 &29 &-0.220(I)     & 0.448     \\\hline
     18 &   40 &-0.132(I)     & 0.273 &  13 &30 &-0.206$\pm$0.041  & 0.458    \\\hline
     18 &   41 &-0.136$\pm$0.020  & 0.284 &  13 &31 &-0.182$\pm$0.036  & 0.445    \\
\hline \hline
 					   
 \end{tabular}				   
  \end{table}				   					   
\clearpage				   
    \begin{table}
    \centering
    \begin{tabular}{cccccccc} 
    & & & 1000 A MeV & & & & I= \mbox{Interpolated}\\  
\hline 
\hline   
    Z &   A &     Mean velocity(cm/ns)  & R.M.S. (cm/ns)    &      Z &   A &	  Mean velocity(cm/ns)    &   R.M.S. (cm/ns) \\\hline\hline
   26& 51 &-0.086$\pm$0.022 &  0.144 &22 &48 &-0.095(I)	 &0.176 	       \\\hline
   26& 52 &-0.075$\pm$0.019 &  0.117 &22 &49 &-0.083$\pm$0.017   &0.162 	   \\\hline
   26& 53 &-0.056$\pm$0.014 &  0.099 &22 &50 &-0.073$\pm$0.015   &0.146 	   \\\hline
   26& 54 &-0.041$\pm$0.010 &  0.066 &22 &51 &-0.074$\pm$0.015   &0.145 	    \\\hline
   26& 55 &-0.031$\pm$0.008 &  0.056 &22 &52 &-0.064$\pm$0.013   &0.126 	   \\\hline
   25& 48 &-0.097$\pm$0.024 &  0.182 &21 &40 &-0.166$\pm$0.033   &0.274 	   \\
\hline
    \hline

 \end{tabular}
  \end{table}
  
\clearpage
					   
    \begin{table}
    \centering
    \begin{tabular}{cccccccc} 
    & & & 1000 A MeV & & & & I= \mbox{Interpolated}\\  
\hline 
\hline   
    Z &   A &     Mean velocity(cm/ns)  & R.M.S. (cm/ns)    &      Z &   A &	  Mean velocity(cm/ns)    &   R.M.S. (cm/ns) \\\hline\hline
   18& 35& -0.181$\pm$0.036  & 0.33926 & 14 & 30& -0.207$\pm$0.031 &  0.405		\\\hline
   18& 36& -0.184$\pm$0.028  & 0.31726 & 14 &31 &-0.203$\pm$0.030  & 0.390	    \\\hline
   18& 37& -0.179$\pm$0.027  & 0.31126 & 14 &32 &-0.192$\pm$0.029  & 0.382	    \\\hline
   18& 38& -0.164$\pm$0.025  & 0.29026 & 14 &33 &-0.191$\pm$0.029  & 0.361	     \\\hline
   18& 39& -0.158(I)	 & 0.28226 & 13 &24 &-0.169$\pm$0.042  & 0.624	    \\\hline
   18& 40& -0.142$\pm$0.021  & 0.26625 & 13 &25 &-0.188$\pm$0.038  & 0.517	    \\
\hline
   
 \hline

 \end{tabular}
  \end{table}

\clearpage
					   
    \begin{table}
    \centering
    \begin{tabular}{cccccccc} 
    & & & 750 A MeV & & & & I= \mbox{Interpolated}\\  
\hline 
\hline   
    Z &   A &     Mean velocity(cm/ns)  & R.M.S. (cm/ns)    &      Z &   A &	  Mean velocity(cm/ns)    &   R.M.S. (cm/ns) \\\hline\hline
   27& 54& -0.060$\pm$0.015  & 0.096 &22& 47 &-0.122(I)	& 0.199 	       \\\hline
   27& 55& -0.058$\pm$0.015  & 0.073 &22& 48 &-0.106(I)	& 0.174 	   \\\hline
   27& 56& -0.049$\pm$0.012  & 0.058 &22& 49 &-0.101$\pm$0.020  & 0.169 	   \\\hline
   26& 51& -0.083$\pm$0.021  & 0.135 &22& 50 &-0.084$\pm$0.017  & 0.150 	    \\\hline
   26& 52& -0.053$\pm$0.013  & 0.118 &22& 51 &-0.089$\pm$0.018  & 0.144 	   \\\hline
   26& 53& -0.050$\pm$0.013  & 0.102 &22& 52 &-0.083$\pm$0.025  & 0.130 	   \\
\hline
    \hline

 \end{tabular}
  \end{table}
  
\clearpage
					   
    \begin{table}
    \centering
    \begin{tabular}{cccccccc} 
    & & & 750 A MeV & & & & I= \mbox{Interpolated}\\  
\hline 
\hline   
    Z &   A &     Mean velocity(cm/ns)  & R.M.S. (cm/ns)    &      Z &   A &	  Mean velocity(cm/ns)    &   R.M.S. (cm/ns) \\\hline\hline
   18 &39& -0.169(I)	 & 0.285&13&  31& -0.228$\pm$0.057   &0.399 	      \\\hline
   18 &40& -0.181$\pm$0.027  & 0.260&12&  24& -0.140$\pm$0.072   &0.511 	  \\\hline
   18 &41& -0.179$\pm$0.027  & 0.250&12&  25& -0.226$\pm$0.056   &0.470 	  \\\hline
   18 &42& -0.161$\pm$0.024  & 0.249&12&  26& -0.230(I)	 &0.468 	   \\\hline
   18 &43& -0.162$\pm$0.032  & 0.242&12&  27& -0.259$\pm$0.065   &0.458 	  \\\hline
   17 &34& -0.152$\pm$0.045  & 0.358&12&  28& -0.214$\pm$0.064   &0.449 	  \\
\hline
   
 \hline

 \end{tabular}
  \end{table}

\clearpage
    \begin{table}
    \centering
    \begin{tabular}{cccccccc} 
    & & & 500 A MeV & & & & I= \mbox{Interpolated}\\  
\hline 
\hline   
    Z &   A &     Mean velocity(cm/ns)  & R.M.S. (cm/ns)    &      Z &   A &	  Mean velocity(cm/ns)    &   R.M.S. (cm/ns) \\\hline\hline
  27 &53& -0.097$\pm$0.013 &  0.106 &22& 47 &-0.156(I)	&0.193		   \\\hline
  27 &54& -0.088$\pm$0.012 &  0.090 &22& 48 &-0.149(I)	&0.174	       \\\hline
  27 &55& -0.073$\pm$0.012 &  0.067 &22& 49 &-0.106$\pm$0.021   &0.169	       \\\hline
  27 &56& -0.074$\pm$0.015 &  0.058 &22& 50 &-0.092$\pm$0.018   &0.153		\\\hline
  26 &51& -0.076$\pm$0.019 &  0.146 &22& 51 &-0.090$\pm$0.018   &0.149	       \\\hline
  26 &52& -0.066$\pm$0.016 &  0.121 &22& 52 &-0.081$\pm$0.016   &0.131	       \\
\hline
   
 \hline

 \end{tabular}
  \end{table}

\clearpage

    \begin{table}
    \centering
    \begin{tabular}{cccccccc} 
    & & & 500 A MeV & & & & I= \mbox{Interpolated}\\  
\hline 
\hline   
    Z &   A &     Mean velocity(cm/ns)  & R.M.S. (cm/ns)    &      Z &   A &	  Mean velocity(cm/ns)    &   R.M.S. (cm/ns) \\\hline\hline
   18& 37& -0.208$\pm$0.031  & 0.313&13 &26& -0.258$\pm$0.052 &  0.522	  	     \\\hline
   18& 38& -0.205$\pm$0.031  & 0.295&13 &27& -0.234$\pm$0.047 &  0.456	  	 \\\hline
   18& 39& -0.200(I)	 & 0.283&13 &28& -0.230(I)    &  0.421	  	 \\\hline
   18& 40& -0.196(I)	 & 0.251&13 &29& -0.228(I)    &  0.407	  	  \\\hline
   18& 41& -0.190$\pm$0.028  & 0.274&13 &30& -0.223$\pm$0.045 &  0.472	  	 \\\hline
   18& 42& -0.151$\pm$0.023  & 0.253&12 &24& -0.287$\pm$0.057 &  0.550	  	 \\
\hline
    \hline

 \end{tabular}
  \end{table}

\clearpage
 
    \begin{table}
    \centering
    \begin{tabular}{cccccccc} 
    & & & 300 A MeV & & & & I= \mbox{Interpolated}\\  
\hline 
\hline   
    Z &   A &     Mean velocity(cm/ns)  & R.M.S. (cm/ns)    &      Z &   A &	  Mean velocity(cm/ns)    &   R.M.S. (cm/ns) \\\hline\hline
  27 &54& -0.069$\pm$0.015 &  0.084 &23& 54 &-0.210(I)	&0.250  	   \\\hline
  27 &55& -0.060$\pm$0.012 &  0.065 &22& 43 &-0.213$\pm$0.043   &0.210         \\\hline
  27 &56& -0.051$\pm$0.011 &  0.058 &22& 44 &-0.206$\pm$0.041   &0.224         \\\hline
  26 &51& -0.110$\pm$0.023 &  0.136 &22& 45 &-0.186$\pm$0.037   &0.209  	\\\hline
  26 &52& -0.082$\pm$0.021 &  0.115 &22& 46 &-0.178$\pm$0.036   &0.195         \\\hline
  26 &53& -0.071$\pm$0.018 &  0.106 &22& 47 &-0.135(I)	&0.194         \\
\hline
  \hline

 \end{tabular}
  \end{table}

\clearpage
	
     \begin{table}
    \centering
    \begin{tabular}{cccccccc} 
    & & & 300 A MeV & & & & I= \mbox{Interpolated}\\  
\hline 
\hline   
    Z &   A &     Mean velocity(cm/ns)  & R.M.S. (cm/ns)    &      Z &   A &	  Mean velocity(cm/ns)    &   R.M.S. (cm/ns) \\\hline\hline
  19& 42& -0.212(I)	& 0.243 &15 &35 &-0.262$\pm$0.039 &  0.321		  \\\hline
  19& 43& -0.210$\pm$0.031  & 0.242 &14 &28 &-0.308$\pm$0.062 &  0.413	      \\\hline
  19& 44& -0.184$\pm$0.037  & 0.217 &14 &29 &-0.305$\pm$0.046 &  0.389	      \\\hline
  19& 45& -0.209$\pm$0.042  & 0.201 &14 &30 &-0.300(I)    &  0.378	       \\\hline
  18& 36& -0.261$\pm$0.039  & 0.314 &14 &31 &-0.295(I)    &  0.319	      \\\hline
  18& 37& -0.255$\pm$0.038  & 0.297 &14 &32 &-0.287$\pm$0.064 &  0.392	      \\
\hline
  \hline

 \end{tabular}
  \end{table}

\clearpage